\documentclass[showpacs,prl,onecolumn,aps,superscriptaddress,preprintnumbers,nofootinbib]{revtex4}
\usepackage[T1]{fontenc}
\usepackage[latin9]{inputenc}
\usepackage{amsmath,amssymb,bigints}
\usepackage{epsfig}
\usepackage{graphicx}
\usepackage{amsmath}
\usepackage{amsfonts}
\def\slashchar#1{\setbox0=\hbox{$#1$}     		
   \dimen0=\wd0                                 	
   \setbox1=\hbox{/} \dimen1=\wd1               	
   \ifdim\dimen0>\dimen1                        	
      \rlap{\hbox to \dimen0{\hfil/\hfil}}      	
      #1                                        	
   \else                                        	
      \rlap{\hbox to \dimen1{\hfil$#1$\hfil}}   	
      /                                         	
   \fi}

\renewcommand{\vec}{\boldsymbol}
\newcommand{\beq}{\begin{equation}}
\newcommand{\eeq}{\end{equation}}
\newcommand{\bea}{\begin{eqnarray}}
\newcommand{\eea}{\end{eqnarray}}
\newcommand{\baa}{\begin{array}}
\newcommand{\eaa}{\end{array}}

\def\eq#1{{Eq.~(\ref{#1})}}
\def\fig#1{{Fig.~\ref{#1}}}

\newcommand{\bas}{\bar{\alpha}_S}

\newcommand{\nn}{\nonumber}

\newcommand{\h}{\frac{1}{2}}

\newcommand{\Lb}{\left(}
\newcommand{\Rb}{\right)}
\def\pom{{I\!\!P}}

\renewcommand{\vec}[1]{\boldsymbol{#1}}

\def\pom{{I\!\!P}}

\begin{document}
\title{CGC/saturation approach: re-visiting the problem of odd harmonics in angular correlations }           
\author{ E.~ Gotsman,}
\email{gotsman@post.tau.ac.il}
\affiliation{Department of Particle Physics, School of Physics and Astronomy,
Raymond and Beverly Sackler
 Faculty of Exact Science, Tel Aviv University, Tel Aviv, 69978, Israel}
 \author{ E.~ Levin}
 \email{leving@post.tau.ac.il, eugeny.levin@usm.cl} \affiliation{Department of Particle Physics, School of Physics and Astronomy,
Raymond and Beverly Sackler
 Faculty of Exact Science, Tel Aviv University, Tel Aviv, 69978, Israel} 
 \affiliation{Departemento de F\'isica, Universidad T\'ecnica Federico Santa Mar\'ia, and Centro Cient\'ifico-\\
Tecnol\'ogico de Valpara\'iso, Avda. Espana 1680, Casilla 110-V, Valpara\'iso, Chile} 

\date{\today}


\begin{abstract}
  In this paper we demonstrate that the selection of 
 events with different multiplicities of produced particles,
 leads to the violation of the azimuthal angular symmetry, $\phi \to \pi - 
\phi$.
 We find  for  LHC and lower energies,
  that this violation can be so large for the events
 with multiplicities $n \geq 2 \bar{n}$, where $\bar{n}$ is
 the mean multiplicity, that it
leads to almost
  no suppression of $v_n$, with odd $n$. 
    However,  this  can only occur if the typical size of the 
dipole in DIS
 with a nuclear target is small, or $Q^2 \,>\,Q^2_s\Lb A; Y_{\rm 
min},b\Rb$,
 where $Q_s$ is the saturation momentum of the nucleus at $Y = Y_{\rm min}$.
     In the case of large sizes of  dipoles, when
 $Q^2 \,<\,Q^2_s\Lb A; Y_{\rm min},b\Rb$,  we show
 that $v_n =0$ for odd $n$.  Hadron-nucleus scattering is discussed.
 
    \end{abstract}

\preprint{TAUP - 3031/18}

\maketitle

\tableofcontents

\flushbottom

\section{Introduction}
In this paper we continue to discuss the azimuthal long range  rapidity
 correlations.
 These correlations were measured in all reactions: hadron-hadron,
 hadron-nucleus and
 nucleus-nucleus scattering,  and they have similar features independent
 of the
 reactions\cite{STAR,PHENIX,PHOBOS,STAR1,CMS,CMS1,CMS2,CMS3,CMS4,ALICE,ALICE1,
ALICE2,ALICE3,ATLAS,ATLAS1,ATLAS2,ATLAS3}. Such  similarity in energy,
 multiplicity and
 transverse momentum dependence as well as in the values of the harmonics
 $v_n$, calls for
 a general explanation. We  believe that the  source of these
 correlations is  the 
Bose-Einstein
 enhancement for  identical gluons.  The origin does not depend on
 the type of the
 reaction, and we have demonstrated that this mechanism alone is able
  to  describe all
 the experimental data\cite{
GLP,GOLE1,GOLE2}. However,  in the effective theory of   high energy
 QCD: CGC/saturation approach (see Ref.\cite{KOLEB} for a review)  
 the resulting 
 angular correlation leads to $v_n$ =0 for all odd $n$ \cite{KOLU1,
KOWE,KOLUCOR,GOLE3,KOLULAST}
 (see also  Refs.\cite{RAJUREV,KOLUREV}). This stems from the 
 symmetry $\phi \to \pi - \phi$, 
where $\phi$ is the azimuthal angle   which is implicitly contained in 
the 
 CGC/saturation approach. This  symmetry 
 does not result from  any fundamental principle,
  and only arises  in the leading order of the approach. Several efforts to
 calculate corrections to the leading order CGC/saturation approach
 have been made
 (see Refs.\cite{KOLULAST,GOLE1,KOSK} which demonstrated that this
 correction violates the symmetry
 which  lead to a $v_n$ for odd $n$.  If these corrections originate from
 the next-to-leading order
 corrections, they should have a parametrically strong suppression, while
 experimentally
 $v_3 < v_2$ but $v_3 \approx v_4$. 
 If we believe that the CGC/saturation
 approach in leading order describes all other physical observables, then
 we interpret  the experimental results, as an indication that  the
 suppression of $v_3$ is of a  numerical nature.

The main idea of this paper is that   selection   by the  multiplicity of
 the event, destroys this 
symmetry, 
and leads to $v_n \neq 0$ for odd $n$.  We consider  the deep inelastic
 scattering on  nuclei,
 which has the most solid theoretical description in the framework of the
 CGC/saturation appproach,
 and allows one to investigate the dependence of the effect on the size
 of the interacting dipoles.

The main result of this paper is that, by  selecting  the event with
 given multiplicity, leads to a
  strong violation of the symmetry in the leading order of  the
 CGC/saturation estimates. We
 show that the violation of this symmetry depends crucially  on
 the sizes of the interacting 
dipoles.
 For DIS with virtuality of photon $Q$, we have two distinct regions.
 For $Q^2_s\Lb A, Y_{\rm min}; b\Rb/Q^2 \,\leq\, 1$  (where $Q^2_s$ 
denotes the saturation scale  and $Y_{min}$ is the minimal value of the rapidity at which we can use CGC approach), it turns out that
 the violation of
 the symmetry is so strong for the events with multiplicities
 $n \,\geq 2 \bar{n}$
\footnote{$Q_s \Lb A, Y_{\rm min}; b\Rb$ is the saturation moment
 of the nucleus at
 low energies ( minimal rapidity $Y_{\rm min}$) at fixed impact parameter
 $b$. $\bar{n}$
 denotes the mean multiplicity in the reaction at fixed rapidity $Y$.},
 that there is practically
 no  suppression of the values of $v_n$ at odd $n$. However, on the other
 hand for 
 $Q^2_s\Lb A, Y_{\rm min}; b\Rb/Q^2 \,\geq\, 2$, we have to deal with the
 violation of
 the geometric scaling behaviour of the scattering amplitude in the
 saturation domain,
 this results in the restoration of the $\phi \to \pi - \phi$ symmetry,
 for the events
 with multiplicities $n \geq 2 \bar{n}$.

In the next section we discuss the origin of the azimuthal angular 
symmetry,  $\phi \to \pi - \phi$ ,
  for the dilute-dilute 
 parton systems scattering in the entire inclusive measurements, and 
 show that this symmetry
 stems from the mixture of events with low multiplicity: multiplicity
 which is less than the
 average multiplicity $\bar{n}$,  and  events with high multiplicity,
 more or equal to
 $2 \bar{n}$. In section 3 we discuss  angular correlations in a  $1 +
 1$ dimensional toy
 model, which can be considered as a theory which describes the 
interaction between QCD
  partons of the fixed sizes. We demonstrate, that in this model,
 $\phi \to \pi - \phi$
 symmetry is reproduced for the entire inclusive measurement. However,
 the selection of
  events with fixed multiplicity violates this symmetry. Our estimates
 shows,  this
 violation is so strong, that for the measurement of the events with
 multiplicities $n
 \geq 2 \bar{n}$ ,  $\bar{n}$ denotes the mean multiplicity in the
 process, does not 
  lead to the suppression of $v_n$ for  odd $n$. In section 4 we consider the
 CGC/saturation approach with a simplified model for the BFKL kernel. For this
 approach we develop a procedure to calculate   the double inclusive 
cross
 section for two gluon production for the events with different 
multiplicities. 

 For  LHC  or lower  energies, in the kinematic
 region where $Q^2_s\Lb A, Y_{\rm min}; b\Rb/Q^2 \,\leq\, 1$
  and for  events with multiplicities $n \,\geq 2 \bar{n}$,  our estimates
 result in a small
   enhancing factor  for $v_n$ with even $n$, and a
damping factor for $v_n$ with odd $n$.
  However, we show that for
 $Q^2_s\Lb A, Y_{\rm min}; b\Rb/Q^2 \,\geq\, 2$
we face a problem of the violation of the geometric scaling behaviour of 
the
 scattering
 amplitude in the saturation domain, which leads to the restoration of the
 $\phi \to 
\pi = \phi$ symmetry in the events with $n \geq 2 \bar{n}$.
   In  the Conclusions we summarize our results.

\begin{figure}[ht]
   \centering
  \leavevmode
      \includegraphics[width=15cm]{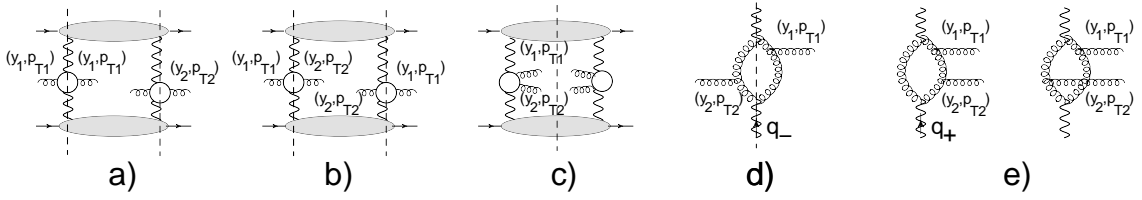}  
  \caption{ Mueller diagrams\cite{MUDI} for the  angular correlations 
for dilute-dilute
 system scattering ( DIS on  a proton target).\fig{dildil}-a: double
 inclusive production
 from two parton showers(two cut BFKL Pomerons\cite{BFKL}).
 \fig{dildil}-b(interference
 diagram): the Bose-Einstein correlations of two identical 
gluons from two parton  
 showers (two cut BFKL Pomerons). \fig{dildil}-c: central
 diffraction production of two gluons in 
the colorless state.\fig{dildil}-d: the structure of the 
vertex for emission of two identical gluons
 in the interference diagram.\fig{dildil}-e: the vertex for
 emission of two gluons in the colorless
 state in  central diffractive production. Wavy lines denote
 the BFKL Pomerons.  The vertical
 dotted lines indicate the final state  that  is measured by
 the detectors.The wavy lines with a
 vertical dotted line, denote the cut Pomerons which corresponds
 to the gluons produced in a
 one parton shower. Its structure is shown in \fig{cutpom}. 
     Helical lines describe gluons. The blobs correspond to
 the amplitude for
 Pomeron-hadron( $\gamma^*$)  scattering, which is integrated over
 the energy
 ($N(Q_T), N_{\gamma^*}\Lb Q, Q_T\Rb$). This integral depends only on the
 transverse momentum of the Pomeron ($Q_T$).
  }
\label{dildil}
 \end{figure}

 \begin{boldmath}
 \section{The dilute-dilute system scattering:
 $\phi \to \pi - \phi$ symmetry and its violation}
 \end{boldmath}
The long range  rapidity correlation for the dilute-dilute system 
scattering (DIS on a proton
 target) stems from the two parton shower production, and can be
 described by the Mueller diagrams 
shown in \fig{dildil}.  These diagrams give the following expression
 for the double inclusive
 cross section for the diagram of \fig{dildil}-a:
  \bea \label{VG1}
 \frac{d \sigma (\fig{dildil}-a)}{d y_1 \, d^2 p_{1T}\, d y_2 
\, d^2 p_{2T} }\, &\propto &\,
\int d^2 Q_T\,N_{\gamma^*}\Lb Q, Q_T\Rb\,N\Lb Q_T\Rb\,\\&\times&\frac{\bas}{p^2_{1T}}\int
 d^2 k_T  \, \phi^{\rm BFKL}\Lb Y - y_1,k_T,\vec{Q}_T\Rb\,\frac{\Gamma_\mu\Lb k_T, p_{1T}\Rb\,
 \Gamma_{\mu}\Lb  k_T, p_{1T}\Rb}{k^2_T\, \Lb \vec{k}_T - \vec{p}_{1T}\Rb^2}\phi^{\rm BFKL}\Lb  y_1,k_T, \vec{Q}_T\Rb \nn\\
 &\times &\,\, \frac{\bas}{p^2_{2T}} \int d^2 l_T  \, \phi^{\rm BFKL}\Lb Y - y_2,
 l_T,-\vec{Q}_T\Rb\frac{\Gamma_\mu\Lb l_T, p_{2T}\Rb\, \Gamma_{\mu}\Lb  l_T,
 p_{2T}\Rb}{l^2_T\, \Lb \vec{l}_T - \vec{p}_{2,T}\Rb^2}\,\phi^{\rm BFKL}\Lb 
 y_1,l_T,-\vec{Q}_T\Rb\,\,\nn
 \eea
 
  In \eq{VG1} $\phi$  at $Q_T=0$ is the solution of the BFKL equation
   \beq \label{BFKL}
\frac{\partial \phi^{\rm BFKL}\Lb y, \vec{k}_T \Rb}{\partial y}\,=\,\bas
 \int \frac{d^2 k'_T}{ \pi}\,\frac{1}{\Lb \vec{k}_T - \vec{k'}_{T}\Rb^2}
\,\phi^{\rm BFKL}\Lb y, \vec{k'}_T\Rb\,\,-\,\,2  \omega_G\Lb  \vec{k}_T\Rb
\,\phi^{\rm BFKL}\Lb y, \vec{k}_T\Rb\,;\eeq
where
\beq \label{OMEGA}
\omega_G\Lb\vec{k}_T\Rb = \h \bas k^2_T \int \frac{d^2 k'_T}{2 \pi}
 \frac{1}{k'^2_T\,\Lb \vec{k}_T - \vec{k'}_{T}\Rb^2 }\,=\,  \bas k^2_T 
\int \frac{d^2 k'_T}{2 \pi} \frac{1}{\Lb k'^2_T\,+\,\Lb \vec{k}_T - \vec{k'}_{T}\Rb^2\Rb\,\Lb 
\vec{k}_T - \vec{k'}_{T}\Rb^2}
\eeq
For $Q_T \neq 0$ the expressions for $\phi$  appear a bit more
  complicated, however, we do 
not
 need to know them, as the $Q_T$ dependance of the BFKL equation
 is determined by the size
 of the largest interacting dipoles. In \fig{dildil} these sizes
 are of the order of the sizes
 of hadrons, which are much larger that $1/p_{iT}$. Therefore, we
 can neglect $Q_T$ in comparison
 with $p_{iT}$ and $k_T$ or $l_T$, which are of the order of $p_{iT}$. 

The diagram of \fig{dildil}-a generates the rapidity correlations,
 but not  correlations in the
 azimuthal angle. The latter stem from two different sources: the
 Bose-Einstein correlations of
 the identical gluons, given by the diagram of \fig{dildil}-b; and
 the central diffractive production
 of two gluons in a colourless state (see \fig{dildil}-c). Both,
  have similar expressions. 
For \fig{dildil}-b we have
  \bea \label{VG2}
 \frac{d \sigma (\fig{dildil}-b)}{d y_1 \, d^2 p_{1T}\, d y_2 \, d^2 p_{2T} }\, &\propto &\,
\frac{1}{N^2_c - 1}\int d^2 Q_T\,N_{\gamma^*}\Lb Q,  Q_T\Rb\,N\Lb \vec{Q}_T +
 \vec{q}_{-}\Rb\,\\&\times&\frac{\bas}{p^2_{1T}}\int d^2 k_T  \, \phi^{\rm BFKL}\Lb Y -
 y_1,k_T,\vec{Q}_T\Rb\,\frac{\Gamma_\mu\Lb k_T, p_{1T}\Rb\, \Gamma_{\mu}\Lb 
 k_T, p_{2T}\Rb}{k^2_T\, \Lb \vec{k}_T - \vec{p}_{1T}\Rb^2}\phi^{\rm BFKL}\Lb  y_2,k_T, \vec{Q}_T\Rb \nn\\
 &\times &\,\, \frac{\bas}{p^2_{2T}} \int d^2 l_T  \, \phi^{\rm BFKL}\Lb Y -
 y_1, l_T,-\vec{Q}_T\Rb\frac{\Gamma_\mu\Lb l_T, p_{1T}\Rb\, \Gamma_{\mu}\Lb 
 l_T, p_{2T}\Rb}{l^2_T\, \Lb \vec{l}_T - \vec{p}_{2,T}\Rb^2}\,\phi^{\rm BFKL}\Lb
  y_2,l_T,-\vec{Q}_T\Rb\,\,\nn
 \eea
 with $\vec{q}_{-}\,=\,\vec{p}_{1T} - \vec{p}_{2T}$. 
  For small
 $y_{12} = y_1 - y_2$ ($\bas y_{12}\,\,\ll\,\,1$) the arguments
 of $\phi$'s in
 both equations are the same, and the correlation function has the form:
 \beq \label{CBE}
 C^{\mbox{\tiny BE}}\Lb L_c | q_{-,T}|\Rb\,\,=\,\,\frac{1}{N^2_c - 1}\frac{\int d^2 Q_T N_{\gamma^*}\Lb Q, Q_T\Rb \,N\Lb \vec{Q}_T + \vec{q}_{-,T}\Rb}{\int d^2 Q_T N\Lb Q_T\Rb \,N\Lb Q_T\Rb}
 \eeq

\begin{figure}[h]
   \centering
  \leavevmode
           \includegraphics[width=10cm]{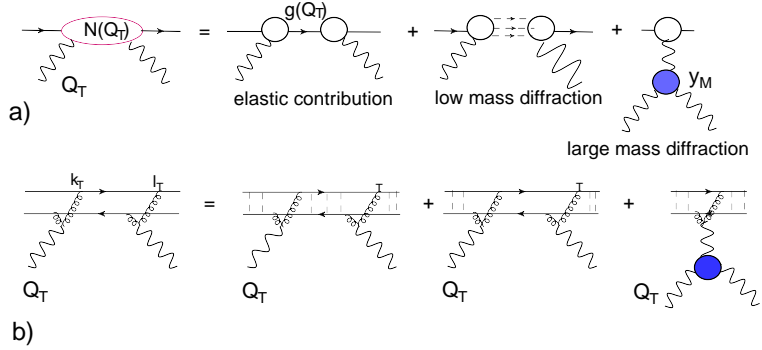}
     \caption{ The graphic form of \protect\eq{N}  in DIS (\fig{n}-a)
 and of $N)Q_T)$ for the
 proton. Large blue blob shows the triple Pomeron vertex. The wavy lines
 describe the BFKL Pomerons.
  }
\label{n}
 \end{figure}

 From \eq{CBE} one can see that the correlation length $L_c$ is determined
 by the dimensional scales
 of the amplitude $N$.    We have  two distinct scales in this amplitude,
 which can be seen from the 
following expression (see \fig{n} and Ref.\cite{GOLE3}):
 
 \bea
 \label{N}
 &&N_{\gamma^*}\Lb Q,  Q_T\Rb \,=\nn\\
 &&\,\, \int d^2 k_T\,d^2 l_T   I_P\Lb \vec{k}_T,\vec{l}_T, - \vec{l}_T + \vec{Q}_T,-\vec{k}_T+  \vec{Q}_T\Rb  \,+\,
 \int_{y_M \gg 1} d y_M \phi^{\rm BFKL}\Lb y_M ,Q,Q_T=0\Rb \,G_{3\pom}\Lb Q_s\Lb y_M\Rb, Q_T\Rb\nn\\
 && N\Lb Q_T\Rb \,=\,\int d M^2 \Bigg(\sum_{i=1}^{M_0} G^2 _i\Lb Q_T\Rb\,\delta\Lb M^2 - M^2_i\Rb \,+ 
 \phi^{\rm BFKL}\Lb y_M \geq y_{M_0} ,k_T,Q_T=0\Rb \,G_{3\pom}\Lb Q_s\Lb y_M\Rb, Q_T\Rb\Bigg)
 \eea
 where 
 \bea 
 F\Lb Q_T\Rb\,\,&=&\,\,\int d^2 r\,d z \, \,e^{i \h  \vec{Q}_T\cdot \vec{r}}\,| \Psi_{\gamma^*}\Lb Q,z, r\Rb|^2\nn \\
  I_P\Lb \vec{k}_T,\vec{l}_T, - \vec{l}_T + \vec{Q}_T,-\vec{k}_T+  \vec{Q}_T\Rb &=& 
  1\,+\, F\Lb 2 \vec{Q}_T\Rb\,+\, F\Lb 2 (\vec{k}_T + \vec{l}_T)\Rb\,+\,  F\Lb 2 (\vec{k}_T - \vec{l}_T - \vec{Q}_T)\Rb \nn\\
  &-&  F\Lb 2 \vec{k}_T \Rb \,-\, F\Lb 2 (\vec{k}_T - \vec{Q}_T)\Rb \,-\,F\Lb 2 \vec{l}_T \Rb \,-\,  F\Lb 2 (\vec{l}_T + \vec{Q}_T)\Rb  \label{IF2} \eea
 
 and $\phi^{\rm BFKL}\Lb y_M ,Q,Q_T=0\Rb$ denotes the unintegrated
 gluon structure function that 
describes
 the BFKL evolution from the  transverse momentum  $Q_s\Lb y_M\Rb$ to $Q^2$.
  The dependence of $G_{3 \pom}$ on $Q_T$ has been discussed in
 Ref.\cite{GOLE3}. 
 The sum over resonance contributions leads  to a scale of about
 the size of the hadron, while  
 the triple Pomeron contribution  for a rapidity $y_M = \ln\Lb 
M^2/M^2_0\Rb$,
 generates a scale which is of the
 order of the saturation scale.

\begin{figure}[h]
   \centering
  \leavevmode
      \includegraphics[width=6.5cm]{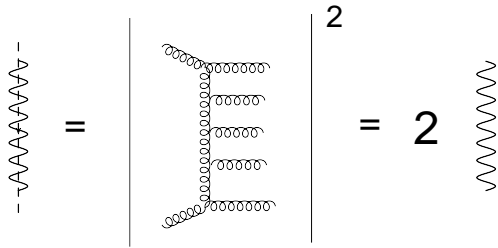}  
  \caption{ The graphic form of  the unitarity constraint (see \eq{UNIT}).
  }
\label{cutpom}
 \end{figure}

  \eq{CBE} does not have a symmetry for $\phi \to \pi - \phi$, and generates
  harmonics $v_n$ both 
with even
 and odd $n$.
  The Mueller diagram of \fig{dildil}-b describes  the interference between
 two produced parton
 showers, since the cut Pomeron is related to the production of the single
 parton shower, as 
 shown in \fig{cutpom}.  From the unitarity constraint 
  \beq \label{UNIT}
 2 \,\mbox{Im} A_{el}\Lb s, b; r \Rb\,\,=\,\,\underbrace{| A_{el}\Lb s, b; r
 \Rb|^2 }_{\rm elastic \,\,cross\,\,section}\,\,+\,\,\underbrace{G\Lb s, b,r\Rb}_{\rm
inelastic\,processes}\,\,\to\,\,2 \,\mbox{Im} G^{\rm BFKL}_\pom\Lb s, b; r \Rb\,=\,G\Lb s, b,r\Rb 
 \,=\,\mbox{cut Pomeron}
  \eeq  
  since the contributions of $| A_{el}\Lb s, b; r\Rb|^2 | $,
 in 
the leading log(1/x) approximation of perturbative QCD  (LL(1/x)A), it 
turns 
out to be negligibly small. 
   
   \eq{CBE} describes the correlations that stem from the event with
 large multiplicity
 of the order of  2\,$\bar{n}$, where   $\bar{n}$ denotes the multiplicity
 of the one parton shower. 
  We need to add the emission of two gluons in the colorless state, 
produced
 in the central
 diffraction process shown in \fig{dildil}-c.
  Generally speaking, the BFKL Pomerons in this diagram are different
 from those in
 \fig{dildil}-b, since the momenta transferred by these Pomerons  have
  longitudinal
 components $Q_{L}$ and $Q_{L} - q_{+,L}$. However, in the leading order
 they can be neglected. 
The vertex for the production of two gluons  (see \fig{dildil}-e),  turns
 out to be twice larger
 (see appendix B in Ref.\cite{GOLE1}) than the  vertex of the gluon emission
 (see \fig{dildil}-d).
  This results in the same contribution of this diagram, as of the diagrams
 of \fig{dildil}-b
with the only difference:  the BFKL Pomeron carry momenta $\vec{Q}_T$ and $
 \vec{Q}_T
 + \vec{q}_{+,T}$ where $\vec{q}_{+,T} 
= \vec{p}_{1T} + \vec{p}_{2T}$. Hence this diagram generates the correlation
 function
 which is equal to
 \beq \label{CCD}
 C^{\mbox{\tiny CD}}\Lb L_c | q_{+,T}|\Rb\,\,=\,\,\frac{1}{N^2_c - 1}\frac{\int d^2 Q
_T N_{\gamma}\Lb Q, Q_T\Rb \,N\Lb \vec{Q}_T + \vec{q}_{+,T}\Rb}{\int d^2 Q_T N\Lb Q_T\Rb \,N\Lb Q_T\Rb}
 \eeq
  
 The sum $ C^{\mbox{\tiny CD}}\Lb L_c | q_{+,T}|\Rb\,+\, C^{\mbox{\tiny 
BE}}\Lb L_c
 | q_{-,T}|\Rb$ has the  symmetry $\phi \to \pi - \phi$. It should
 be emphasized that this
 symmetry is a direct consequence of an entirely inclusive measurement,
without any selection of the event accordingly to  multiplicity. 

However, one can see that this symmetry stems from the mixture of
 two  events with quite
 different multiplicities: 
diagrams of \fig{dildil}-a and \fig{dildil}-b describe the events
 with the multiplicity
 $n = 2\bar{n}$, while the diagram of \fig{dildil}-correspond to the
 events with low
 multiplicities $n \ll \bar{n}$. In other words, if we select events
 with large multiplicities
 so that $n \geq 2 \bar{n}$,  we have no
 $\phi \to \pi - \phi$ symmetry, and
 the source for the azimuthal angular correlation is the Bose-Einstein
 enhancement.

\begin{figure}[ht]
   \centering
  \leavevmode
      \includegraphics[width=12cm]{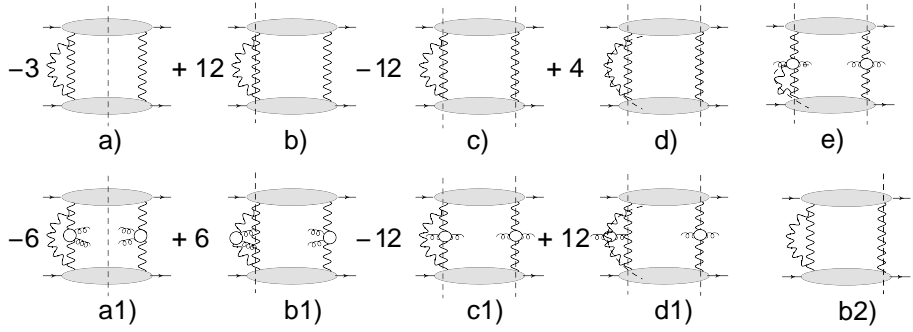}  
  \caption{AGK \cite{AGK} cutting rules for the exchange of three Pomerons
 (\fig{endil}-a - \fig{endil}-d) 
  and Mueller diagrams for the central diffractive production of two gluons
 (\fig{endil}-a1 and \fig{endil}-b1) and for two particle correlations
 (\fig{endil}-c1 and \fig{endil}-d1). \fig{endil}-e is the diagram that
 can violate the AGK cutting rules. However, this diagram accounts for the 
 Pomeron loop of the size of $Y/2$, where $Y$ is the total rapidity. 
Hence,
  the contribution of this diagram   turns out to be much smaller than
 the contributions of the diagrams \fig{endil}-a1 and \fig{endil}-b1. 
The 
notations are the
 same as in \fig{dildil}.
  }
\label{endil}
 \end{figure}

 It is instructive to note, that   for the entire inclusive
 measurement, this symmetry  is
 not  violated in the next to leading approximation. For a dilute-dilute
 system the first
 corrections are related to accounting for  the one Pomeron loop (see
 \fig{endil}).
 In inclusive measurements, we  take into account the processes of two 
gluon diffractive production given by \fig{endil}-a1 and \fig{endil}-b1. 
\fig{endil}-a1 describes the process of central diffractive production 
with low multiplicity, while \fig{endil}-b1 shows the diffractive 
production
 which is accompanied by the multi-gluon generation from the one parton 
shower,
 with multiplicity $\bar{n}$.  In \fig{endil}-c1 and \fig{endil}-d1 the double
 inclusive cross sections are shown for the event with multiplicities
 $2 \bar{n}$
(\fig{endil}-c1) and $ 3 \bar{n}$ (\fig{endil}-d1).
\fig{endil}-a - \fig{endil}-d demonstrate the AGK cutting rules and 
provide
 the weight
 of the processes with different multiplicities: $n \ll\bar{n}, \bar{n},
 2 \bar{n}$
 and $3 \bar{n}$, respectively. Taking into account the simple
  combinatorics for
 two gluon diffractive  emission, and the emissions from the different
 parton showers shown in \fig{endil}-a1 - \fig{endil}-d1, one can see,
 that the double gluon cross sections  and the central diffractive
 contributions, are the same as for the emission of the two gluon showers.
 In other words, in  the next-to-leading order diagrams, the contributions
 with different multiplicities are canceled, leading  to vanishing 
contributions for  inclusive measurements. We postpone the calculation 
of the
 combinatoric coefficient  to the next section, but we would
 like to note that  central diffraction can come from the diagram of
 \fig{endil}-b1, but it cannot  originate from the diagram of 
\fig{endil}-b2.
 Hence, the symmetry $\phi \to \pi - \phi$ is not violated in the
 next-to-leading order.
 It shows that the symmetry $\phi \to \pi - \phi$ is an inherent
 feature of QCD, at least in the leading log(1/x) approximation.

 The contribution of the first Pomeron loop is well known,
and its calculation leads to  lengthy and cumbersome formulae, which
 can be
 found in Refs.\cite{BRN,LELULOOP,LMP,ACKLLS,AKLL}.  Our strategy is to clarify
 all essential points using the simplified version of the Pomeron calculus in
 1+1 space-time, which we discuss in the next section.

 Prior to doing  so, we wish  to comment on the AGK cutting rules in QCD.
The  AGK cutting rules have been discussed  and proven in 
Refs.\cite{AGK1,AGK2,AGK3,AGK4,AGK5,AGK6,AGK7} for the inclusive cross sections.
  In Ref.\cite{AGK8} it
 is shown that the AGK cutting rules are violated for  double inclusive
 production. This violation is intimately related to the enhanced diagrams
 \cite{AGK7,AGK8}, and reflects the fact that different cuts of the triple BFKL 
Pomeron
 vertex,  lead to different contributions,  as can be seen from 
\fig{dildil} and
 \fig{endil}. We will  not consider such diagrams. In principle, we 
can consider 
 diagrams of the type of \fig{endil}-e, however, these diagrams  
 correspond to the contribution of the small Pomeron loop($\sim Y/2$ ,where $Y$
 is the total rapidity). 
 Hence, their contributions are  small compared to the
 diagrams \fig{endil}-a1 and \fig{endil}-b1.

 \section{ The BFKL Pomeron calculus  in zero transverse dimensions:
 correlations in  hadron-nucleus scattering}
 
 \subsection{Generalities}
 In this section we consider a simplified model for the Pomeron
 interaction, in which we neglect  the fact that this interaction can 
change
 the sizes of  dipoles\cite{MUDIP,LELU}. In such an approach the 
DIS process with a nucleus target  appears to be the same as 
 proton-nucleus scattering. In this model  the scattering amplitude
 ($N$) is a function of one variable: $Y$ for which we have a
 simplified Balitsky-Kovchegov equation\cite{BK} of the form:
 \beq \label{TM1}
 \frac{d \,N\Lb Y\Rb}{d \,Y}\,\,=\,\,\Delta\Lb N\Lb Y\Rb\,\,-
\,\,N^2\Lb Y \Rb\Rb
 \eeq
 The solution to this equation has the form
 \beq \label{TM2}
 N\Lb Y\Rb\,\,=\,\,\frac{\gamma\,e^{\Delta\,Y}}{1 + \gamma \Lb e^{\Delta Y} - 1\Rb}\,\,=\,\,\frac{\gamma \,z}{1 + \gamma\Lb z - 1\Rb}
 \eeq
 where $N\Lb Y = 0 \Rb\,=\,\gamma$ and $ z = e^{\Delta Y}$.  
  In the linear approximation, when $N^2 \ll N$ \eq{TM2}
 degenerates to 
 \beq \label{TM3}
  \frac{d \,N\Lb Y\Rb}{d \,Y}\,\,=\,\,\Delta\, N\Lb Y\Rb
  \eeq
  hence, $\Delta $ is the intercept of the BFKL Pomeron.
  
 \begin{figure}[ht]
   \centering
  \leavevmode
      \includegraphics[width=12cm]{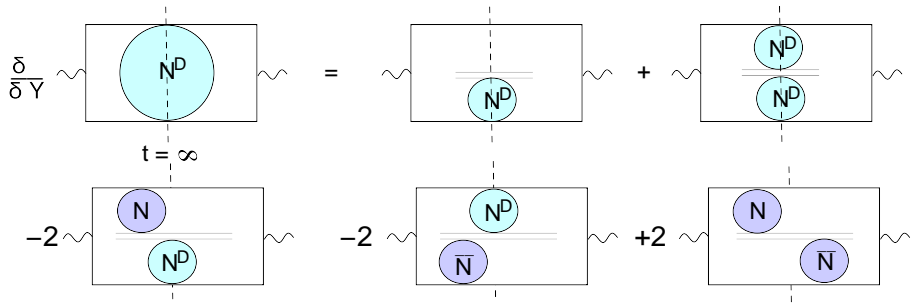}  
  \caption{The graphic form of the equation of Ref.\cite{KOLE}
  for  diffractive production ($N^D$)  in the BFKL Pomeron
 calculus,  in zero transverse dimensions. $N$ and $\bar{N}$ denote
 the elastic amplitudes  with initial conditions: $N\Lb Y=0\Rb =\gamma$
 and $\bar{N}\Lb Y=0\Rb = \bar{\gamma}$.
  }
\label{eqkl}
 \end{figure}

\begin{figure}[ht]
   \centering
  \leavevmode
      \includegraphics[width=9cm]{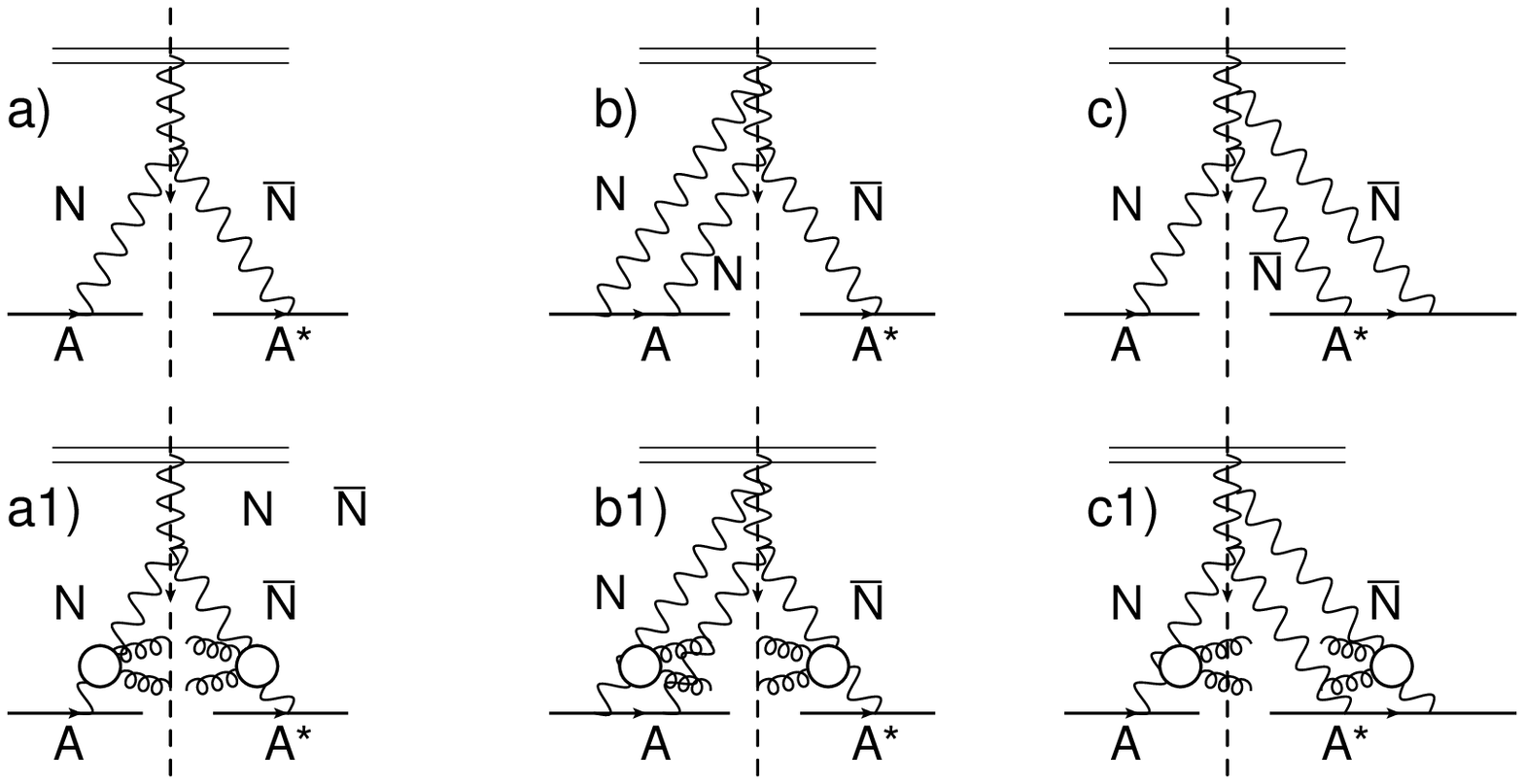}  
  \caption{The first diagrams for $N^D$ (\fig{eldd}-a - \fig{eldd}-c) and
 for the two gluon central diffraction productions (\fig{eldd}-a1 - 
\fig{eldd}-c1).}
\label{eldd}
 \end{figure}

  The equation for the process of the diffractive dissociation which
 was proven in QCD \cite{KOLE}, transforms into the following equation
 in the framework of the BFKL Pomeron calculus in  zero transverse
 dimension\cite{KOLE,BGLM,BKKMR}:
  \beq \label{TM3}
 \frac{d  N^D \Lb  Y, Y_{min}\Rb}{d \Delta\,Y} \, =\, N^D \Lb 
 Y, Y_{min}\Rb \,+\,\Lb N^D \Lb  Y, Y_{min}\Rb\Rb^2\,-\,2\,\Big( N\Lb
 Y\Rb\,+\, \bar{N}\Lb Y\Rb\,\Big) \,  N^D \Lb  Y, Y_{min}\Rb\,+\,2 \,
  N\Lb Y\Rb \,  \bar{N}\Lb Y\Rb 
 \eeq
 $N^D$ denotes the cross section for diffractive production with a
 rapidity gap larger than      $Y_{min}$. Generally speaking this cross
 section can be
 viewed as a product of the amplitude $A$ and the complex conjugate
 amplitude $A^*$. $N$  and $\bar{N}$ are  the amplitudes for elastic
 scattering in $A$ and $A^*$,respectively. \fig{eldd} illustrates this
 notation. From this figure one can see,  the difference between $N$
 and $\bar{N}$.    For the calculation of the 
processes of 
 central diffractive production, we only need  to separate the 
diagrams of \fig{endil}-b2 from the other diagrams,
 which do not contribute to the diffraction.
 
 The solution to \eq{TM3}  takes the following form
 
 \beq \label{TM4}
 N^D \Lb  Y, Y_{min}\Rb\,\,=\,\,\frac{\gamma\,z}{1+ \gamma\Lb z - 1\Rb} \,+\, \frac{\bar{\gamma}\,z}{1+ \bar{\gamma}\Lb z - 1\Rb}   
  \,-\,\frac{\gamma + \bar{\gamma} \,+\,2\,\gamma\,\bar{\gamma}\Lb z_{min} - 1\Rb}{  1 - \gamma\,\bar{\gamma}\Lb z_{min} - 1\Rb^2 \,+\, 
 \Lb  \gamma + \bar{\gamma} \,+\,2\,\gamma\,\bar{\gamma}\Lb z_{min} - 1\Rb \Rb\,\Lb z - 1\Rb}
 \eeq 
 where $z \,=\,e^{\Delta\, Y}$ and $z_{min} \,=\,e^{\Delta\,Y_{min}}$.
 \eq{TM4} reduces to a  more transparent expression for
 $\gamma = \bar{\gamma}$ :
 
 \beq \label{TM5}
  N^D \Lb  Y, Y_{min}\Rb\,\,=\,\,\frac{2\,\gamma\,z}{1+ \gamma\Lb z - 1\Rb} \,\,-\,\,\frac{2 \gamma \,z}{1 \,+\,\gamma\Lb 2 z - z_{min} - 1\Rb}
  \eeq

  For $z_{min} = 1$, \eq{TM4} and \eq{TM5} give the total cross section 
 for diffraction production, which has the form:
   \beq \label{TM6}
 N^D \Lb  Y, Y_{min}\Rb\,\,=\,\,\frac{\gamma\,z}{1+ \gamma\Lb z - 1\Rb} \,+\, \frac{\bar{\gamma}\,z}{1+ \bar{\gamma}\Lb z - 1\Rb}   
  \,-\,\frac{\gamma + \bar{\gamma} }{  1  \,+\, 
 \Lb  \gamma + \bar{\gamma} \Rb\,\Lb z - 1\Rb}
 \eeq   
Using \eq{TM6} we can calculate the central diffraction production
 cross section, which is equal to
\beq \label{TM7}
\sigma_{\rm CD}\,\,=\,\,\Gamma^2\Lb 2 \pom \to  2 G\Rb \,
\,\gamma\frac{\partial}{\partial \gamma} \, \bar{\gamma}
\frac{\partial}{\partial \bar{\gamma} } N^D \Lb  Y,
 Y_{min}\Rb\Big{|}_{\gamma = \bar{\gamma}}\,\,=\,
\,\Gamma^2\Lb 2 \pom \to  2 G\Rb\,\frac{2\gamma^2
 \,z\,\Lb z - 1\Rb}{\Lb 1 + 2\gamma\Lb z - 1\Rb\Rb^3}
\eeq
where $\Gamma^2\Lb 2 \pom \to  2 G\Rb$ denotes the vertex  of two
 gluon production from  Pomeron exchange.
  \subsection{Healing the Finkelstein -Kajantie  disease}
\begin{figure}[ht]
   \centering
  \leavevmode     
      \includegraphics[width=3cm]{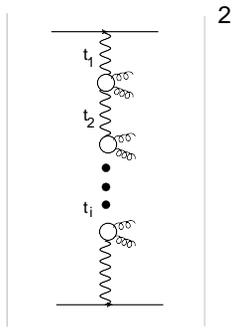}  
  \caption{The process of  multi central diffractive production 
due to multi Pomeron  exchanges. The wavy lines describe the 
Pomerons.}
\label{mpom}
 \end{figure}

    Having calculated the central diffractive production, we can
 shed light on an old problem which was understood in the 1960's:
   the process of production  of pairs of the gluons separated by
 a large rapidity gap (LRG), could  violate  
 unitarity constraints. Indeed, 
   even if the resulting Green function of the Pomeron  produces an
 amplitude that does not depend on the energy of the multi-Pomeron 
exchange,
 shown in \fig{mpom}, and leads to the power-like increase of the 
scattering
 amplitude \cite{KTM}(see also Refs.\cite{FK,ANSW,KMRFK}). This phenomenon
 was unfairly called   the Finkelstein -Kajantie  disease (see review of
 Ref.\cite{ANSW}). The widely held  opinion at that time was that
 $\Gamma^2\Lb 2 \pom \to 2G\Rb \propto t_i$. Such suppression, turns
 out to be sufficient to suppress this process at high energies. However,
  no reason for such a behavior has been found over almost five decades,
 and as we have argued, no such suppression appears in QCD for two gluon
 production by the Pomeron.

    On the other hand,  one can see that \eq{TM7} leads to the cross
 section for central diffraction, which decreases as $e^{- \Delta Y}$, 
without any problem with  unitarity.   The 
emission of more than one pair of
 gluons, which is equal to
    
    \beq \label{FK1}
\sigma^{(k)}_{\rm CD}\,\,=\,\,\Lb \Gamma^2\Lb 2 \pom \to  2 G\Rb\Rb^k \,\Lb \,\gamma\frac{\partial}{\partial \gamma} \, \bar{\gamma}\frac{\partial}{\partial \bar{\gamma} }\Rb^k N^D \Lb  Y, Y_{min}\Rb\Big{|}_{\gamma = \bar{\gamma}}\,\,\xrightarrow{z\,\,\gg\,1}\,\frac{1}{z} \,\,=\,\,e^{- \Delta\,Y}
\eeq    
   where $k$ which denotes the number of produced pairs of gluons, 
 does not change the behavior of the amplitude at large values of $Y$.
    
    We need to compare this behavior of the scattering amplitude  
 with
 the contribution of the diagram of \fig{mpom}, which leads to 
    \beq \label{FK2}  
     A\,\propto \,e^{\Delta_{\rm sum}\,Y}; ~~~~\mbox{with}~~~~~~~\Delta_{\rm sum}    \,=\,\Delta \,+\, \Gamma^2\Lb 2 \pom \to  2 G\Rb
    \eeq    
    
    In other words, the `fan' diagrams of \fig{eldd} generate the survival
 probability, which suppress both the power-like growth of the `bare' 
 Pomeron, and
    the increase due to the multi-Pomeron production.
    
    Generally speaking, we showed the suppression in a rather specific 
model,
 but one can see that the amplitude $N^D \to 1$ at large $Y$,  and it
 approaches this limit as $e^{-\Delta y}$:  $N^D = 1 - {\cal O}\Lb 
 e^{-\Delta Y}\Rb$.   After  differentiation over $\gamma$
 and $\bar{\gamma}$, only the correction term  remains.  We will
 see below that this structure is  preserved in  QCD.
  \subsection{Generating functional for the production processes.}
  \eq{FK2} resolves the FK problem, but it also  shows that contribution 
of
  central diffraction turns out to be rather small. In other words, 
if we suggest an experiment to measure the events with multiplicity $n
 \geq \bar{n}$, we expect, at high energies,  the violation of $\phi \to 
\pi - \phi$ symmetry,
 to be small.
  Fortunately, for the  Balitsky-Kovchegov cascade, we know how to 
calculate
 the events with different multiplicities  in the BFKL Pomeron calculus in
 zero transverse dimension \cite{LEPR}. To do   this, we need 
to introduce
 the generating function\footnote{In the general case of the BFKL Pomeron
 calculus in  four dimensions  this function will be a
 functional\cite{MUDIP,LELU}.}
  \beq \label{GF1}
   Z\Lb w,\bar{w},v; Y\Rb\,\,=\,\,\sum^{\infty}_{k=0,l=0,m=0}\,P\Lb k,l,m;Y\Rb w^k\,\bar{w}^l\,v^m
   \eeq
   where $k$($l$) denotes the number of uncut Pomerons in the amplitude
 and in the complex conjugate  amplitude, and $m$ is the number of
 cut Pomerons at rapidity $Y$.  In Ref.\cite{LEPR}  it is shown that
 this generating function satisfies the following equation:
   \bea \label{GFEQ}
 &&  \frac{\partial   Z\Lb w,\bar{w},v; Y\Rb}{\partial\,\Delta\,Y}\,\,=\,\, \\
 && -\Bigg(  w (1 - w)  \frac{\partial   Z\Lb w,\bar{w},v; Y\Rb}{\partial\,w}\,+\, \bar{w} (1 - \bar{w})  \frac{\partial   Z\Lb w,\bar{w},v; Y\Rb}{\partial\,\bar{w}}  \,+\,\Lb 2 w \bar{w} - 2 ( w + \bar{w})  v + v^2 + v\Rb  \frac{\partial   Z\Lb w,\bar{w},v; Y\Rb}{\partial\,v} \Bigg)\nn
   \eea
   The solution to this equation takes the following form
   \beq \label{GFZ}
     Z\Lb w,\bar{w},v; Y\Rb\,\,=\,\,\frac{ w}{(1 - w)
 (z - 1)\,+\,1}\,+\,\frac{ \bar{w}}{(1 - \bar{w}) (z - 1)\,+\,1} 
   \,-\,\frac{ w +\bar{w}  - v}{(1 - w - \bar{w} + v)(z - 1) +1
}     \eeq
    where $z = e^{\Delta Y}$.
    
    We can identify the scattering amplitude with $N\Lb \gamma,\bar{\gamma},
\gamma_{\rm in};Y\Rb\,=\,1 -  Z\Lb 1 - \gamma, 1 - \bar{\gamma},1 - 
\gamma_{\rm in}; Y\Rb  $\cite{BK,LELU,LEPR} and obtain the following
 expression for the amplitude:
   \beq \label{GFN}
   N\Lb \gamma,\bar{\gamma},\gamma_{\rm in};Y\Rb \,\,=\,\,\,\,\frac{
 \gamma z}{\gamma (z - 1)\,+\,1}\,+\,\frac{ \bar{\gamma} z}{\bar{\gamma}
 (z - 1)\,+\,1}    \,-\,\frac{ \Lb \gamma + \bar{\gamma}  - \gamma_{\rm in
 }\Rb z} {\Lb \gamma + \bar{\gamma}  - \gamma_{\rm in}\Rb(z - 1) +1
}      
 \eeq   
   where $\gamma = \bar{\gamma}$ denotes the amplitude for the elastic
  interaction of a single dipole with the target at $Y=Y_0$, while
 $\gamma_{in}$ denotes the amplitude of the inelastic interaction. Due to
 the AGK cutting rules, $\gamma_{in} = 2 \gamma = 2 \bar{\gamma}$.
   
   Note that \eq{GFN} leads to \eq{TM6} for the
 total cross section of diffraction production at $\gamma_{\rm
 in} = 0$. This condition means that we do not produce even one cut  Pomeron.

   From \eq{GFN} we can calculate the result for the total inclusive
 measurement. Indeed, the total cross section for central
 diffraction, without any selection with respect of the multiplicity
 of the events, is equal to   
   \beq \label{GFCD}
   \sigma_{\rm CD}\,\,=\,\,\Gamma^2\Lb 2 \pom \to  2 G\Rb \,\,\gamma
\frac{\partial}{\partial \gamma} \, \bar{\gamma}\frac{\partial}{\partial
 \bar{\gamma} }  N\Lb \gamma,\bar{\gamma},\gamma_{\rm in};Y\Rb\Big{|}_{
\gamma_{\rm in} = 2 \gamma =  2\bar{\gamma}}\,\,=\,\,\Gamma^2\Lb 2 \pom
 \to  2 G\Rb\,2 \,\gamma^2\,z\,\Lb z - 1\Rb
   \eeq
The double inclusive cross section for two cut Pomeron production which
 is accompanied by any number cut and uncut Pomerons is equal to
   \beq \label{GFBE}
   \sigma_{\rm BE}\,\,=\,\,\frac{1}{N^2_c - 1}\Gamma^2_G \,\,\gamma^2_{\rm in}\frac{\partial}{\partial \gamma_{\rm in}} \, \frac{\partial}{\partial \gamma_{\rm in} }  N\Lb \gamma,\bar{\gamma},\gamma_{\rm in};Y\Rb\Big{|}_{\gamma_{\rm in} = 2 \gamma =  2\bar{\gamma}}\,\,=\,\,\frac{1}{N^2_c - 1}\Gamma^2_G\,2\,\gamma^2_{\rm in} \,z\,\Lb z - 1\Rb
   \eeq
\eq{GFBE} describe the  Bose-Einstein interference diagram and the
 contribution for the entire inclusive measurement with $\Gamma_G$
 being the Mueller vertex for the inclusive production of one gluon.

One can see that for 
\beq \label{CDBE}
\Gamma^2\Lb 2 \pom \to  2 G\Rb = \frac{4}{N^2_c - 1}\Gamma^2_G
\eeq
$ \,\sigma_{\rm BE} = \sigma_{\rm CD}$, 
which results in the symmetry $\phi \to \pi - \phi$.

The contribution  to the correlation function of the even $n$ with fixed 
multiplicity : $n = k 
\bar{n}$,
 is given by the following formula:
\bea \label{GFFN}
\sigma^{BE}_n\,\,=\,\,\frac{1}{N^2_c - 1}\Gamma^2_G \,
\,\frac{\gamma^k_{\rm in}}{k!}\frac{\partial}{\partial
 \gamma^k_{\rm in}} \,   N\Lb \gamma,\bar{\gamma},
\gamma_{\rm in};Y\Rb\Big{|}_{\gamma_{\rm in} =0, 
\gamma = \bar{\gamma}}\,\,&=&\,\,\frac{1}{N^2_c -
 1}\Gamma^2_G\gamma^k_{\rm in} \frac{z \Lb z - 1\Rb^{k -
 1}}{\Lb1 + 2\,\gamma\,\Lb z - 1\Rb\Rb^{k+1}}\nn\\
&\,=\,&\frac{1}{N^2_c - 1}\Gamma^2_G\Lb 2 \gamma\Rb^k
 \frac{z \Lb z - 1\Rb^{k - 1}}{\Lb1 + 2\,\gamma\,\Lb z
 - 1\Rb\Rb^{k+1}}\eea
The cross section for  central diffraction with the same multiplicity
 of produced gluons takes the form:
\bea \label{GFCDN}
\sigma^{CD}_n\,\,&=&\,\, \Gamma^2\Lb 2 \pom \to  2 G\Rb \,\,\,\gamma 
\frac{\partial}{\partial \gamma} 
\,\bar{\gamma} \frac{\partial}{\partial \bar{\gamma} }\,\frac{\gamma^k_{\rm in}}{k!}\frac{\partial}{\partial \gamma^k_{\rm in}} \,   N\Lb \gamma,\bar{\gamma},\gamma_{\rm in};Y\Rb\Big{|}_{\gamma_{\rm in} =0, \gamma = \bar{\gamma}}\nn\\\,\,&=&\,\,\Lb k + 2\Rb \Lb k + 1\Rb\Gamma^2\Lb 2 \pom \to  2 G\Rb\gamma^2\,\gamma_{\rm in}^k\frac{z \Lb z - 1\Rb^{k+1}}{\Lb 1 + 2 \,\gamma\Lb z - 1\Rb\Rb^{k + 3}}\nn\\
&=&\,\Lb k + 2\Rb \Lb k + 1\Rb\Gamma^2\Lb 2 \pom \to  2 G\Rb\,\gamma^2\,\Lb 2\, \gamma\Rb^k\frac{z \Lb z - 1\Rb^{k+1}}{\Lb 1 + 2 \,\gamma\Lb z - 1\Rb\Rb^{k + 3}}\eea

\begin{figure}[ht]
   \centering
  \leavevmode     
      \includegraphics[width=7cm]{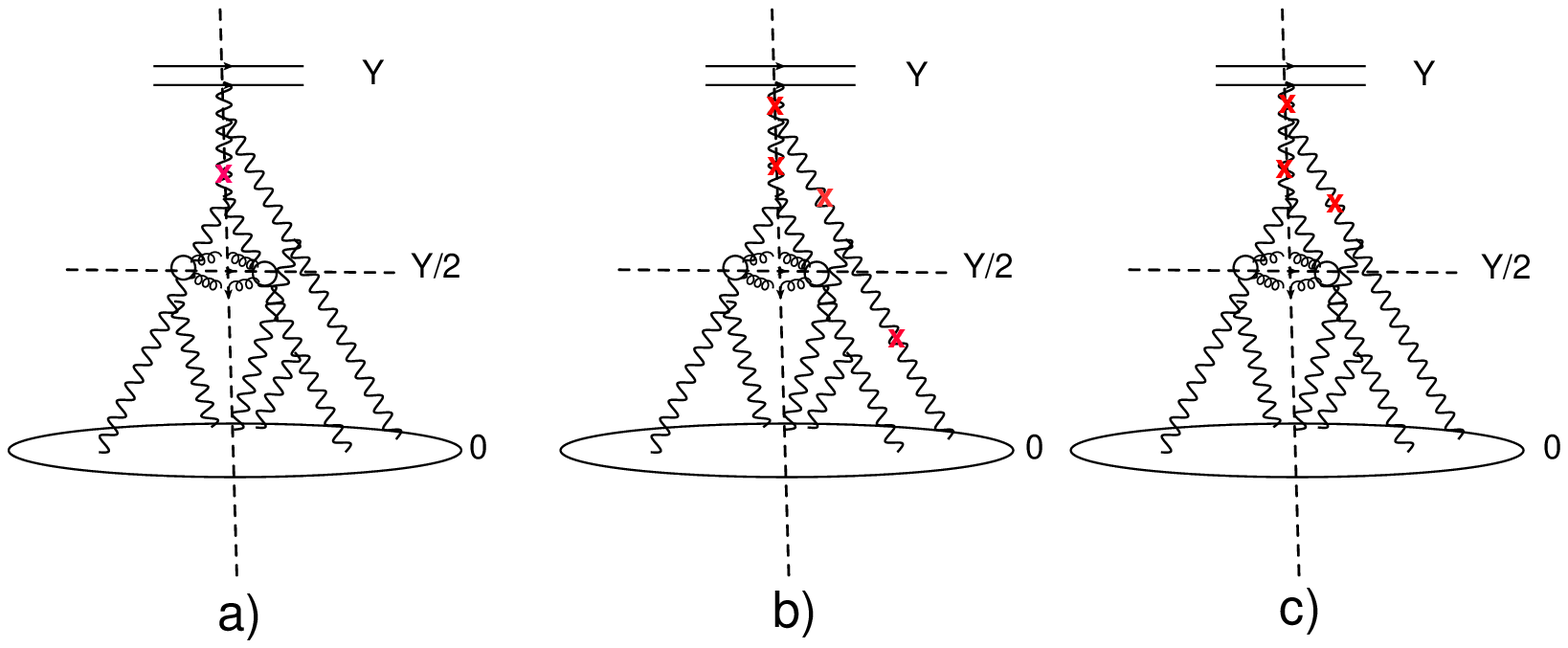}  
  \caption{Graphic forms of $\sigma^{CD}_0$ (\fig{x01}-a) 
 and $\sigma^{CD}_1$ (\fig{x01}-b). The dotted line show the cut Pomerons.}
\label{x01}
 \end{figure}

 However, the simple formulae of \eq{GFFN} and \eq{GFCDN} are only  
correct, 
 if we do not fix the rapidity of the emitted particles. Indeed,
 if the emitted gluons have rapidity $y_1 \approx y_2 = \h Y$, we have
 to calculate  $\sigma^{BE}_n$ and $ \sigma^{CD}_n$ using \eq{GFFN}
 and \eq{GFCDN} for rapidity $\h Y$ and insert in this formulae
 $\gamma =\bar{\gamma} =  \frac{ \gamma z}{\gamma (z - 1)\,+\,1}$
 with $z = \exp\Lb \h \Delta  Y\Rb$( see \fig{x01}-a).  For example
 $\sigma^{CD}_0 $ takes the form
 \bea \label{CDN0}
 \sigma^{CD}_0\,\,&=&\,\, \Gamma^2\Lb 2 \pom \to  2 G\Rb \,\,
\,\gamma \frac{\partial}{\partial \gamma} 
\,\bar{\gamma} \frac{\partial}{\partial \bar{\gamma} }\, \,  
 N\Lb \gamma,\bar{\gamma},\gamma_{\rm in};\h Y\Rb\Bigg{|}_{\gamma_{\rm in} =0,
 \gamma = \bar{\gamma}\,=\, \frac{\gamma \exp\Lb \h \Delta Y\Rb}{  1 + \gamma
 \Lb \exp\Lb \h \Delta Y\Rb - 1\Rb}}\nn\\
&=& 2\Gamma^2\Lb 2 \pom \to  2 G\Rb\,\frac{\Lb \gamma e^{\Delta Y}\Rb^2 \,\Lb
 1 \,\,+\,\,\gamma \,e^{\h \Delta Y}\Rb}{\Lb 1 + \gamma\Lb 2 \,e^{ \Delta Y}
\,-\,e^{\h \Delta Y}\Rb\Rb^3}
\eea
where $\gamma $ denotes the dipole amplitude at $Y=Y_0$, and we assumed
 that $\exp\Lb \h \Delta Y\Rb \,\gg\,1$.

 We  need to find the scattering amplitude with one cut Pomeron, which
 is equal to
 \beq \label{GFN1}
A\Lb\mbox{1 cut Pomeron},Y\Rb\,=\,\gamma_{\rm in}\Lb  \frac{\partial}{\partial\,\gamma_{\rm in}}N\Lb \gamma,\bar{\gamma},\gamma_{\rm in};Y\Rb \Big{|}_{\gamma_{\rm in} =0,\gamma=\bar{\gamma}}\Rb\,=\,\frac{2 \gamma e^{\Delta Y}}{\Lb 1\,+\,2\,\gamma \,\Lb e^{\Delta Y}\,-\,\Rb\Rb^2}
\eeq

 Using \eq{GFN1} we obtain
 \bea \label{CDN1}
  \sigma^{CD}_1\,\,&=&\,\, \Gamma^2\Lb 2 \pom \to  2 G\Rb \,\,\,A\Lb\mbox{1 cut Pomeron},\h Y\Rb \,\frac{\partial}{\partial \gamma_{\rm in}} \Lb \gamma \frac{\partial}{\partial \gamma} 
\,\bar{\gamma} \frac{\partial}{\partial \bar{\gamma} }\, \,   N\Lb \gamma,\bar{\gamma},\gamma_{\rm in};\h Y\Rb\Rb\Bigg{|}_{\gamma_{\rm in} =0, \gamma = \bar{\gamma}\,=\, \frac{\gamma \exp\Lb \h \Delta Y\Rb}{  1 + \gamma \Lb \exp\Lb \h \Delta Y\Rb - 1\Rb}}\nn\\
&=& 12\,\Gamma^2\Lb 2 \pom \to  2 G\Rb\,\frac{\Lb \gamma e^{\Delta Y}\Rb^3 \,\Lb 1 \,\,+\,\,\gamma \,e^{\h \Delta Y}\Rb^2}{\Lb 1\,+\,2 \,\gamma\,e^{\h \Delta Y}\Rb^2\,\Lb 1 + \gamma\Lb 2 \,e^{ \Delta Y}\,-\,e^{\h \Delta Y}\Rb\Rb^4}
\eea 
 
  We also need to take into account the events with multiplicities less
 than $\bar{n}$, which stem from the processes of diffraction 
dissociation.
 For this we need to replace in \eq{CDN1} $A\Lb\mbox{1 cut Pomeron},\h Y\Rb
  $ with the amplitude of the cross section of the diffraction production 
 $N\Lb \gamma, \gamma, 0, Y\Rb$. In \fig{x01}-c we show an example of
 such processes. We denoted the cross section for such processes by
 $\sigma^{\rm CD}_{\h} $.

  Introducing the damping factor $R_1$ as
 \beq \label{RD}
  R_1\Lb \gamma, \Delta Y\Rb\,\,=\,\,\frac{\sigma^{\rm CD}_0 \,+\,\sigma^{\rm CD}_1\,+\,\,\sigma^{\rm CD}_{\h}}{\sum^\infty_{n = 0} \,\sigma^{\rm CD}  }
  \eeq
 Note that the value of $R_1$ depends crucially  on the values of
 the amplitude $\gamma$ and of the Pomeron intercept $\Delta$.  For DIS
 this amplitude is proportional to $\bas^2$ and we expect that it is
 small. The value of $\Delta$ in DIS is a function of the value of $Q$.
 It changes from $\Delta = 0.1$ for $Q \sim 1 \,Gev$ to  $\Delta = 0.3$
 at $Q = 10\,GeV$.   For the estimates in the
 kinematic region  of the LHC, we took  $\Delta = 0.2$. 
 \begin{figure}[ht]
   \centering
  \leavevmode     
      \includegraphics[width=7cm]{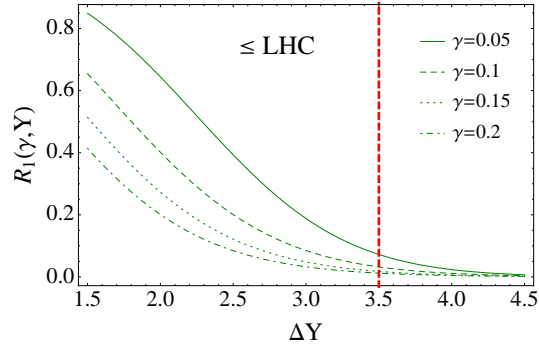}  
  \caption{The damping factor $R_1$  versus $\Delta Y$ for different
 values of the amplitude $\gamma$ at $Y=Y_{\rm min}$}  
  \label{rsw}
 \end{figure}
   
\begin{figure}[ht]
   \centering
  \leavevmode   
  \begin{tabular}{c c c}
      \includegraphics[width=7cm]{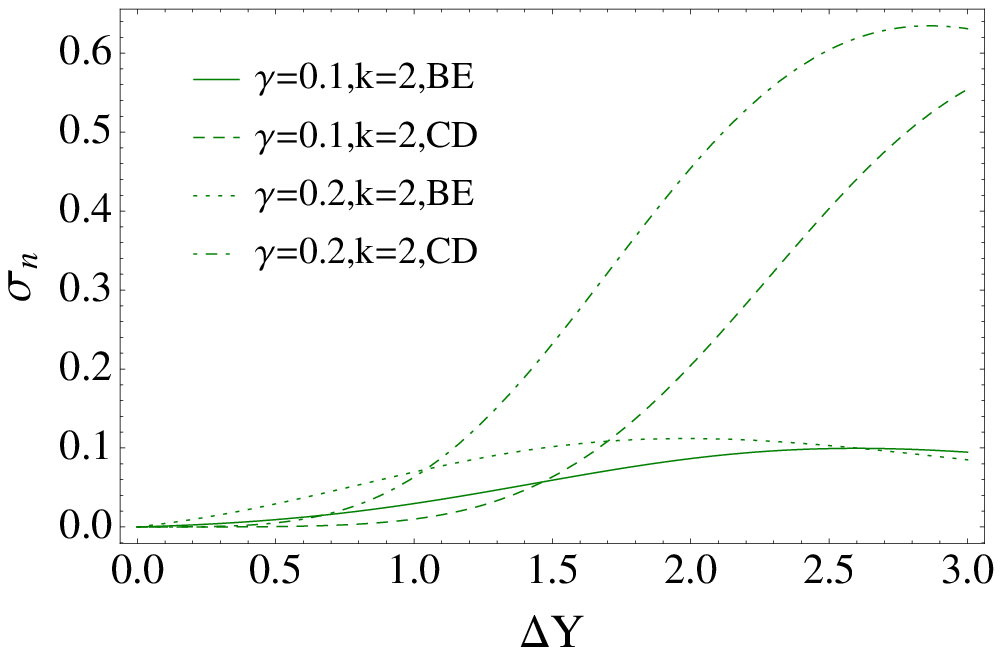}  &~~~~~~&  \includegraphics[width=7cm]{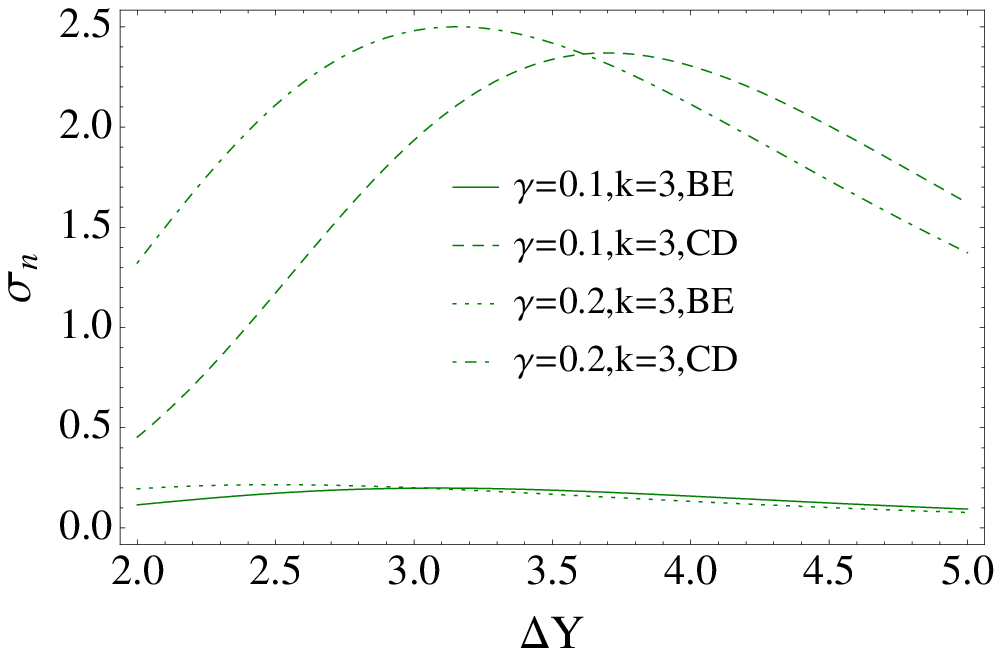}\\
      \fig{becd}-a & & \fig{becd}-b\\
      \end{tabular}
        \caption{$\sigma^{BE}_n$ and $\sigma^{CD}_n$ with the
 multiplicity of the produced gluons equal to $n=k \bar{n}$ versus $Y$.}
\label{becd}
 \end{figure}

  In the next section we consider a more realistic approach to determine
 these parameters.
  In \fig{becd} we compare $\sigma^{BE}_n$ and $\sigma^{CD}_n$  for
 the events with fixed multiplicities: $n\,=\,k \bar{n}$, where $\bar{n}$
 denotes the average multiplicity. One can see that for different values 
of $k$ and at
 different rapidities ,we have different relations between central
 diffraction production and Bose-Einstein enhancement.
   
.    
   
      \begin{boldmath}
  \subsection{ Schwimmer model for hadron-nucleus scattering with induced
 $\phi \to \pi - \phi$ symmetry: $v_{n,n} $ and $v_n$.}
  \end{boldmath}
 In the previous sections we discussed the BFKL Pomeron calculus
 in zero transverse dimension as a simplified model for the QCD
 cascade. However, it was noted  long ago\cite{SCHW} that this type 
of
  model, is the correct approach to the hadron-nucleus interaction in
 the soft Pomeron calculus, with a triple Pomeron interaction. Indeed,
 the soft Pomeron, generally speaking, has 
 a trajectory $\alpha_\pom (t) \,=\,1 + \Delta + \alpha_\pom' t$, but
 $\alpha'_\pom \ln (s/s_0) $     turns out to be smaller than $R^2_A$ ($
 \alpha'_\pom \ln (s/s_0)\,\ll\,R^2_A$ for all accessible energies,
 and, therefore, it can be neglected.  Since the vertex
  $G_{3\pom}$ is small, we can neglect the contribution of the Pomeron
 loops which are proportional to $G^2_{3  \pom}$,  and consider only
 `fan' diagrams (see examples of these diagrams in \fig{eldd}-a -
 \fig{eldd}-c) which are of the order of $(G_{3 \pom}\,g\,S_A\Lb
 b\Rb)^k$, where $k$ denotes the number of  Pomerons, $g$  the
 strength of the vertex of the Pomeron-nucleon interaction and $S_A\Lb 
b\Rb$ is given by
 \beq \label{SM1}
 S_A\Lb b\Rb\,\,=\,\,\int^{+\infty}_{-\infty} \!\!\!\!\!\!d z\,\,
 \rho\Lb \sqrt{z^2+ b^2}\Rb~~~~~~~\mbox{with}~~~~~\int d^2 b\, S_A\Lb b \Rb\,=\,A
 \eeq
  $b$ denotes the impact parameter of the nucleon  and $\rho$  the 
density
 of the nucleons in the nucleus.
 
 Bearing this in mind,  the general \eq{GFN} can be
 re-written replacing $\gamma = \bar{\gamma} $ by $ G_{3 \pom}
 \,g\,S_A\Lb b \Rb/(2 \Delta)$ (see Refs. \cite{BGLM,BKKMR}
\footnote{In these references the Schwimmer approach of Ref.\cite{SCHW}
 was generalized for the intercept of the Pomeron $\Delta_\pom > 0$.}).
We suggest the following strategy to find $G_{3 \pom}$ from the experimental
 data on  soft interactions:
\beq \label{SM2}
G_{3 \pom} \,g \,\,=\,\,\h\,\frac{\sigma^{\rm HM}_{\rm dif}}{\sigma_{el}}
\eeq
In \eq{SM2} we assumed, that both elastic and single diffraction can be
 described as the exchange of two Pomerons, and that the $t$ dependence of
 the triple Pomeron vertex can be neglected in comparison with the elastic
 slope. It should be noted that both assumptions are in agreement with high
 energy phenomenology (see for example Ref.\cite{GLM2CH}). The value of the
 cross section for  diffractive production in the region of high mass
 $\sigma^{\rm HM}_{\rm dif}$ is taken from Ref.\cite{GLM2CH}. From \eq{SM2}
 we see that the parameter $G_{3 \pom} g z$ which enters \eq{GFFN} and
 \eq{GFCDN} can be written as
\beq \label{SM3}
s_{pp}\,\,\equiv\,\,G_{3 \pom} g z\,\,=\,\,\Bigg(\h\,\frac{\sigma^{\rm
 HM}_{\rm dif}}{\sigma_{el}}\bigg) \sigma_{\rm in}
\eeq
where $\sigma_{\rm in}$ denotes the inelastic cross section for 
proton-proton
 interaction at high energy. Considering $z \,\gg\,1$ we can re-write
 \eq{GFFN} and \eq{GFCDN}  in terms of $s_{pp}$ in the form:
\bea 
\sigma^{BE}_n\,\,&=&\frac{1}{N^2_c - 1}\,\Gamma^2_G\frac{\Lb 2 s_{pp}\,S_A\Lb b \Rb \Rb^k }{\Lb 1 + 2\,s_{pp} \,S_A\Lb b \Rb\Rb^{k+1}}; \label{FBE}\\
\sigma^{CD}_n\,\,&=&\frac{\Lb k + 2\Rb \Lb k + 1\Rb}{4}\Gamma^2\Lb 2 \pom \to  2 G\Rb\,\frac{\Lb 2\,s_{pp}\,S_A\Lb b \Rb\Rb^{k+2}}{\Lb 1 + 2 \,s_{pp}\,S_A\Lb b \Rb\Rb^{k + 3}}\label{FCD}
\eea
Using these equations, we can estimate the contribution to the double
 inclusive production of the terms which violate $\phi \to \pi - \phi$
 symmetry, due to selections of the events with restricted multiplicities.
 If we select all events with multiplicity $n \geq 2 \bar{n}$, where
 $\bar{n}$ is the average multiplicity, the central diffraction
 production with multiplicities $n\, < 2 \, \bar{n}$ will be not
 measured and, therefore,  has to be subtracted from the inclusive
 measurements that show the $\phi \to \pi - \phi$    symmetry. We can
 introduce the parameters
\beq \label{SM3}
R_1\Lb W\Rb \,\,=\,\,\frac{\sigma^{\rm CD}_0 \,+\,\sigma^{\rm CD}_1}
{\sum^\infty_{n = 0} \,\sigma^{\rm CD}_n};~~~~~~~~~~~~~~~~~R_0\Lb W\Rb
 \,\,=\,\,\frac{\sigma^{\rm CD}_0 }{\sum^\infty_{n = 0} \,\sigma^{\rm
 CD}_n};
\eeq
whose values show the suppression of the symmetry violation terms,
 with respect to symmetry preserving one. $R_1$ characterizes the
 violation of the symmetry in the measurement with the
 multiplicity $n \,\geq\,2 \bar{n}$, while $R_0$ shows this
 violation for the measurements with large multiplicity $n \geq \bar{n}$.
 \begin{figure}[ht]
   \centering
  \leavevmode   
      \includegraphics[width=7cm]{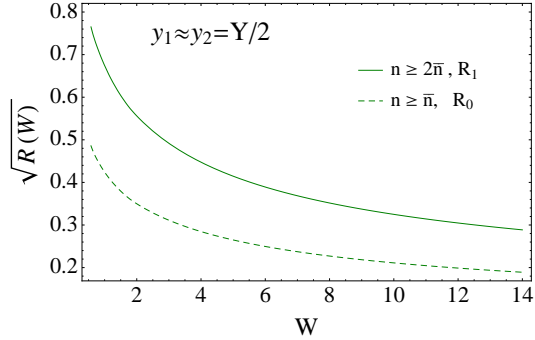}    
        \caption{$R_1$ and $R_0$ versus energy $W$ for proton-gold
 scattering for $y_1 \approx y_2 = \h Y$.}
\label{r}
 \end{figure}
   
 
 \fig{r} shows the value of these parameters in the Schwimmer
 model with the parameters that are found using    \eq{SM1}
 and \eq{SM2}. In \fig{r} we plot the result of the estimates
 from \eq{CDN0} and \eq{CDN1} fixing $y_1 \approx y_2 = \h Y$.
  For making such estimates we parametrized
 \eq{SM2} as $s_{pp}\,\,=\,\,0.025\exp\Lb 0.2\,Y\Rb$.
  
  Using $R$ we can express $v_{n,n}$ and $v_n$ through the 
 harmonics that has been evaluated from the Bose-Einstein 
correlations\cite{
GLP,GOLE1,GOLE2}. Indeed, it turns out that
\beq \label{SMV}
v_{2n, 2n}\,=\,\Lb 2 - R\Rb \,v^{\rm BE}_{2n, 2n};~~~~\,v_{2n-1, 2n-1}\,=\, R \,v^{\rm BE}_{2n-1, 2n-1};~~~~
\,v_{2 n}\,=\,\sqrt{ 2 - R} \,v^{\rm BE}_{2n};\,~~~~v_{2n-1}\,=\, \sqrt{R} \,v^{\rm BE}_{2n-1};
\eeq
 One can see from   \fig{r} ( see  \eq{FBE} and \eq{FCD})  that the odd
 harmonics are small at high energies but they are sizable at $W = 5.5 
TeV$ 
since $\sqrt{R_1} = 0.32$ and $\sqrt{R_0}\,=0.25$ at this energy. Recall,
 that at the moment, this is the highest energy available for 
hadron-nucleus scattering data.
~
      
      ~
      
      ~
  \section{QCD cascade}

  In the previous section we discussed the simplified model  which actually
 reproduces  only  two features of the CGC/saturation  approach: the form of
 \eq{TM1} and the relation given by  \eq{CDBE}. In this section we wish to 
repeat
 the previous estimates on  more general grounds of  high energy 
QCD. 
However, it should be stressed that the longitudinal structure of the QCD
 cascade, is very close to that  of the model, that we have considered.
 In particular, we can introduce the
 generating functional\cite{KLP}
  \beq \label{QCDGF}
   Z\Lb \{w(r_k)\},\{\bar{w}(r_l)\},\{v(r_m)\}; Y\Rb\,\,=\,\,\sum^{\infty}_{k=0,l=0,m=0}\,P^m_{k,l}\Lb r_1,\dots r_k; r_1,\dots,r_l; r_1,\dots ,r_m ;Y\Rb \prod^k_{i=0} w(r_i) \,\,\prod^l_{i=0} \bar{w}(r_i)\,\, \prod^k_{i=0} v(r_i)  
   \eeq
   where $ w(r_i),\bar{w}(r_i)$ and $v(r_i)$ are arbitrary functions
 and $P^m_{k,l} $ is the probability to have $k$ and  $l$ dipoles with
 the coordinates $r_i$ in uncut Pomerons, while $m$ is the  number of  
dipoles in
 the cut Pomerons at rapidity $Y$. For this functional, we can write
 the functional linear equations which are similar to the ones in the 
simplified model (see Ref.\cite{KLP} for details).
   Bearing these general features  in mind, we suggest a simpler approach
 which is based on the non-linear Balitsky-Kovchegov equation\cite{BK}, 
and on  general properties of the solution, that we have discussed above
  in the simplified model.
   
  \subsection{The  simplified non-linear equation}
  The simplified version of the Balitsky-Kovchegov (BK) equation was
 proposed in Ref.\cite{LETU} and has been discussed in detail in
 Ref.\cite{LEPP,KLT}. Here we give a brief review of this equation
 and concentrate our efforts on finding the solution in the form which
 can be used for calculating  the angular correlations. The BK equation
 takes the following form:
  \bea \label{BK}
\displaystyle{\frac{\partial N\Lb Y; \vec{x}_{01}, \vec{b}
 \Rb}{\partial Y}}\,&=&\,\displaystyle{\frac{\bas}{2\,\pi}\int d^2 
 \mathbf{x_{2}}\,K\Lb \vec{x}_{01};\vec{x}_{02},\vec{x}_{12}\Rb\Bigg\{
  N \Lb Y; \vec{x}_{02},\vec{b} - \h \vec{x}_{12}\Rb +  N\Lb Y; 
\vec{x}_{12},\vec{b} - \h \vec{x}_{02}\Rb\,-\,N\Lb Y; \vec{x}_{01
},\vec{b} \Rb}\,\nn\\
&-&\,\displaystyle{ N\Lb Y; \vec{x}_{02},\vec{b} - \h \vec{x}_{12}\Rb\,
 N\Lb Y ;\vec{x}_{12},\vec{b} - \h \vec{x}_{02}\Rb\Bigg\}}
~~~~~~\mbox{where}~~~~ \displaystyle{K\Lb \vec{x}_{01};\vec{x}_{02},
\vec{x}_{12}\Rb}\,\,=\,\,  \displaystyle{\frac{\mathbf{x^2_{01}}}{
\mathbf{x^2_{02}}\,\,
{\mathbf{x^2_{12}}} } }
\eea
where $N\Lb Y; \vec{x}_{01}, \vec{b}\Rb$ denotes the dipole scattering 
amplitude.

Since the analytical solution to \eq{BK} has not been found,
in Ref.\cite{LETU} it was suggested to simplify the kernel by taking into
 account only log contributions.     We have two kinds of logs: $ \Big(\bas
 \ln\Lb x^2_{01}\,\Lambda^2_{QCD}\Rb\Big)^n$ in the perturbative QCD kinematic
 region where  $x^2_{01}\,Q^2_s\Lb Y,b\Rb\,\,\equiv\,\,\tau\,\,\ll\,\,1$; and
 $  \Big(\bas \ln\Lb x^2_{01}\,Q^2_s\Lb Y,b \Rb \Rb\Big)^n$ inside the
 saturation domain ($\tau\,\gg\,1$), where $Q_s\Lb Y, b \Rb$ is the
 saturation scale. To sum these logs we need to modify the BFKL kernel
 in different ways in the two kinematic regions. From the formal point of 
view,
 this simplification means that we consider only the leading twist 
contribution
 to the BFKL kernel,  which  includes all twist contributions in the form 
of \eq{BK}.
For the perturbative QCD region of $\tau \,\ll\,1$, the logs originate 
from
 $ x^2_{02} \sim x^2_{12} \,\ll\, x^2_{01}$ resulting in the following
 form of the kernel $ \displaystyle{K\Lb \vec{x}_{01};\vec{x}_{02},
\vec{x}_{12}\Rb}$ \cite{LETU}
 \beq \label{K1}
 \int d^2 x_{02}\, \displaystyle{K\Lb \vec{x}_{01};\vec{x}_{02},\vec{x}_{12}\Rb}\,\,\,\,\rightarrow\,\pi\, x^2_{01}\,\int^{\frac{1}{\Lambda^2_{QCD}}}_{r^2} \frac{ d x^2_{02}}{x^4_{02}}
\eeq
The non-linear BK equation in this region  can be written as 
\beq \label{BK1}
\frac{\partial^2 n\Lb Y; \vec{x}_{01}, \vec{b}
 \Rb}{\partial Y\,\partial \ln\Lb 1/(x^2_{01}\, \Lambda^2_{QCD})\Rb}\,\,\,=\,\,\frac{\bas}{2}\,\Big( 2 n\Lb Y; \vec{x}_{01}, \vec{b}
 \Rb\,\,- \,n^2 \Lb Y; \vec{x}_{01}, \vec{b}
 \Rb\Big)
\eeq
for $n\Lb Y; \vec{x}_{01}, \vec{b}
 \Rb\,=\,N\Lb Y; \vec{x}_{01}, \vec{b}
 \Rb/x^2_{01}$ .

Inside of the saturation region where $\tau\,\,>\,\,1$ the logs 
   originate from the decay of a large size dipole into one small
 size dipole  and one large size dipole.  However, the size of the
 small dipole is still larger than $1/Q_s$. This observation can be
 translated in the following form of the kernel
\beq \label{K2}
 \int \, \displaystyle{K\Lb \vec{x}_{01};\vec{x}_{02},\vec{x}_{12}\Rb}\,d^2 x_{02} \,\,\rightarrow\,\pi\, \int^{x^2_{01}}_{1/Q^2_s(Y,b)} \frac{ d x^2_{02}}{x_{02}^2}\,\,+\,\,
\pi\, \int^{x^2_{01}}_{1/Q^2_s(Y, b)} \frac{ d |\vec{x}_{01}  - \vec{x}_{02}|^2}{|\vec{x}_{01}  - \vec{x}_{02}|^2}
\eeq

Inside the saturation region the BK equation takes the form
\beq \label{BK2}
\frac{\partial^2 \widetilde{N}\Lb Y; \vec{x}_{01}, \vec{b}\Rb}{ \partial Y\,\partial \ln r^2}\,\,=\,\, \bas \,\left\{ \Lb 1 \,\,-\,\frac{\partial \widetilde{N}\Lb Y; \vec{x}_{01}, \vec{b} \Rb}{\partial  \ln x^2_{01}}\Rb \, \widetilde{N}\Lb Y; \vec{x}_{01}, \vec{b}\Rb\right\}
\eeq
where 
 $\widetilde{N}\Lb Y; \vec{x}_{01}, \vec{b}\Rb\,\,=\,\,\int^{ x^2_{01}} d x^2_{02}\,N\Lb Y; \vec{x}_{02}, \vec{b}\Rb/x^2_{01}$ .

The new kernel in the anomalous dimension representation has the form:
\bea \label{KSM}
\chi\Lb \gamma\Rb\,\,=\,\,\left\{\begin{array}{l}\,\,\,\frac{1}{\gamma}\,\,\,\,\,\mbox{for}\,\,\,\tau\geq \,1\,;\\ \\
\,\,\,\frac{1}{1 \,-\,\gamma}\,\,\,\,\,\mbox{for}\,\,\,\tau\,\leq\,1\,; \end{array}
\right.
\eea
This should be compared with the   full BFKL kernel  in the Mellin 
transform:
\beq \label{KML}
\chi\Lb \gamma\Rb \,\,=\,\,\int \frac{d \xi}{2 \pi i}\,e^{- \gamma \xi}\,\displaystyle{K\Lb \vec{x}_{01};\vec{x}_{02},\vec{x}_{12}\Rb}\\,\,=\,\,
2 \psi(1) \,- \,\psi(\gamma) \,-\,\psi(1 - \gamma)
\eeq
where $\xi \,=\,\ln(x^2_{01}/x^2_{02})$ and $\psi(z) = d\ln \Gamma(z)/d
 z$ with $\Gamma(z)$ equal to Euler gamma function.

One can see that the advantage of the simplified kernel of \eq{KSM} is 
that,  in Double Log Approximation (DLA) for $\tau < 1$,
 it provides a matching with the DGLAP evolution equation\cite{DGLAP}.

  \subsection{Solution}
  
  \subsubsection{Perturbative QCD region ($\tau\,<\,1$)}
For $\tau = x^2_{01}\,Q^2_s(Y, b)\,<\,1$, we can neglect the non-linear 
term
 in \eq{BK1}. The equation leads to the DLA solution that has the form

\beq \label{SOLK1}
N\Lb Y; x_{01}, b \Rb\,\,=\,\,N_0\,\exp\Lb \sqrt{-\,\xi_s\,\xi}\,\,+\,\,\xi\Rb \,\xrightarrow{\tau \to 1;  \zeta \,\to\,0_-} \,\,N_0 e^{\h \zeta }\,\exp\Lb - \frac{\zeta^2}{8 \xi_s}\Rb
\eeq
where we use the following notations:
\beq \label{NOTA}
\xi_s\,\,=\,\,4\,\bas\Lb Y - Y_{\rm min}\Rb;~~~~~\xi\,\,=\,\,\ln\Lb x^2_{01}\, Q^2_s\Lb Y=Y_{\rm min}; b\Rb \Rb ;~~~~~~~~\zeta\,\,=\,\,\xi_s\,+\,\xi;
\eeq
The solution of \eq{SOLK1} provides the boundary condition for the 
solution
 inside the saturation region:
\beq \label{INCK1}
N\Lb Y; \zeta = 0_-(\xi = -\xi_s), b \Rb\,\,=\,\,N_0\Lb b \Rb;~~~~~~~~~~~~ \frac{\partial \ln  N\Lb Y; \zeta=0_- (\xi = -\xi_s), b \Rb}{\partial \zeta}\,\,=\,\,\h;
\eeq

As  was expected, in the vicinity of the saturation scale
 ( $\zeta \ll 8 \xi_s$), the amplitude shows  geometric 
scaling behavior, being a function of  only one variable 
$\zeta$ \cite{GSV},
\beq \label{GSK1}
N\Lb Y; r, b \Rb\,\,\propto \Lb r^2 Q^2_s\Lb Y, b \Rb \Rb^{ 1 - \gamma_{cr}}
\eeq 

where $\gamma_{cr}$ the critical anomalous dimension is equal to $\h$.


  \subsubsection{Saturation region ( $\tau \,> \,1$).}
In this region  we  look for a solution in the form\cite{LETU}
\beq \label{NSAT}
\widetilde{N}\,\,=\,\,\int^{\xi}_{\xi_s}
 d \xi'\,\Big( 1\,-\,e^{ - \phi(\xi',Y)}\Big)
\eeq

Substituting \eq{NSAT} into \eq{BK2} we obtain
\beq \label{NSAT1}
 \phi'_Y\,e^{ - \phi}\,\,=\,\, \bas \widetilde{N}\,e^{ - \phi}
\eeq
Canceling $e^{ - \phi}$ and differentiating with respect to $\xi$ we
  obtain the equation in the form:
\beq \label{EQXIY}
\frac{\partial^2 \phi}{ \partial Y\,\partial \xi}\,\,=\,\,\,\bas\,\Big( 1\,-\,e^{ - \phi\Lb Y;\xi\Rb}\Big)
\eeq
Using variable $\xi_s$ and $\xi$ we can rewrite \eq{NSAT1} in the form
\beq \label{EQ}
\frac{\partial^2 \phi}{ \partial \xi_s\,\partial \xi}\,\,=\,\,\frac{1}{4}\Big( 1\,-\,e^{ - \phi\Lb Y;\xi\Rb}\Big)~~~\mbox{or in the form}~~~~
\frac{\partial^2 \phi}{ \partial  \zeta^2}\,\,-\,\,\frac{\partial^2 \phi}{ \partial  x^2}\,\,=\,\,\frac{1}{4}\Big( 1\,-\,e^{ - \phi\Lb Y;\xi\Rb}\Big)
\eeq
with  $\zeta$ defined in \eq{NOTA} and $x= \,\xi_s - \xi$.

\eq{EQ} has a general traveling wave solution (see Ref.\cite{POL} formula 
{\bf 3.4.1})

\beq \label{GSOL}
\int^\phi_{\phi_0}\frac{d \phi'}{\sqrt{c \,+\,\frac{1}{2 ( \lambda^2 - \kappa^2)}\Big( \phi'  - 1 + e^{-\phi'}\Big)}}\,\,=\,\,  \kappa \,x +\lambda \,\zeta
\eeq
where $c, \phi_0,\lambda$ and $\kappa$ are arbitrary constants that should be
  determined from the initial and boundary conditions.

From the matching with the perturbative QCD region (see \eq{INCK1}) we have
 the following initial conditions for small values of $\phi_0$:
\beq \label{IC}
\phi\Lb t \equiv \zeta  = 0, x\Rb\,\,=\,\,\phi_0\Lb b \Rb\,;\,\,\,\,\,\,\phi'_\zeta\Lb t \equiv \zeta = 0, x\Rb\,\,=\,\,\frac{1}{2}\,\phi_0\Lb b \Rb
\eeq

These conditions allow us to find that $\kappa=0$ and  $c=0$ for
 $\phi_0 \,\ll\,1$.
 Therefore,the  solution of \eq{GSOL} leads to  geometric scaling as
 it depends only on one variable:   $z$.     \,   For small values of 
$\phi_0$, it takes  the
 form\cite{LETU,POL}.
\beq \label{SOLF}
\sqrt{2}\,\int^\phi_{\phi_0}\frac{d \phi'}{\sqrt{ \phi' \,-\,1+ \,e^{-\phi'}}}\,\,=\, \,\zeta
\eeq

  \subsection{Formulation of the problem}

\begin{figure}[ht]
   \centering
  \leavevmode     
      \includegraphics[width=15cm]{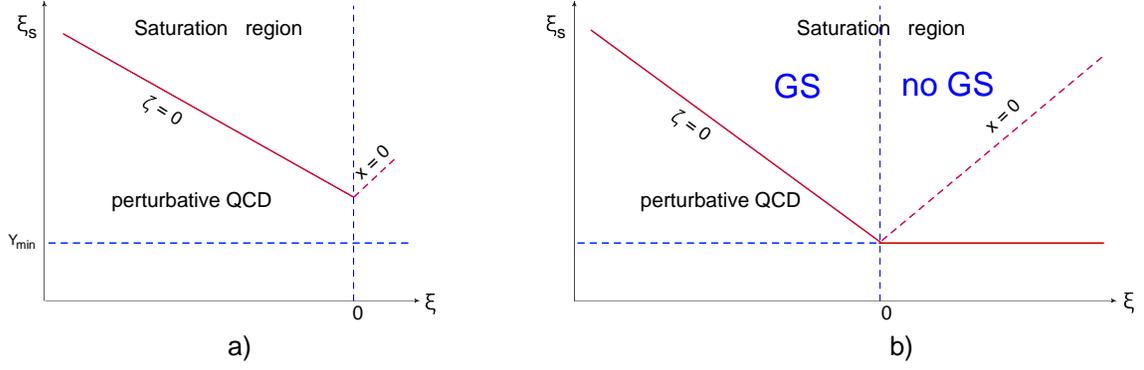}  
  \caption{ QCD map. \fig{map}-a shows the kinematic regions
 for the case when $N\Lb Y = Y_{\rm min},\xi\Rb \,\ll\,1$ while
 in \fig{map}-b we show  the kinematic regions for $ N\Lb
 Y = Y_{\rm min},\xi\Rb \,\sim\,\,1$ (see Ref.\cite{KLT} for
 more details). Note  that the saturation domain in this
 case can be divided in two subregions: (i) for $\xi < 0$,  where
 we expect the geometric scaling behaviour of the scattering
 amplitude, and $\xi > 0$ where there is no such behaviour. }
\label{map}
 \end{figure}

 The previous sections give a brief review of the simple approach
 to the QCD cascade of one dipole which interacts with a target.
 In \fig{map}-a one can see the two distinct kinematic regions which we
 have considered above: the perturbative QCD region with $\tau <1$, and 
the saturation domain for which $\tau > 1$. The key physics idea of
 the description of  DIS with a  target nucleus and/or hadron-nucleus
 collisions in the framework of  the CGC/saturation approach, is that the
 physics in the saturation region is determined by the new dimensional
 parameter: the saturation scale, and if it is a dilute system of 
partons, it
 does not depend on the detailed structure of the projectile.
In DIS we have 
a
 dipole  of  size $r \sim  1 /Q$. For the hadron-nucleus collision 
  we identify the projectile hadron with a dipole of the same size. 
 In these processes we have two different situations which are shown in
 \fig{map}-a and \fig{map}-b. For small dipoles $ \tau_{m}\,\equiv\,r^2 
\,Q^2_s\Lb A; Y_{\rm min};b\Rb\,\ll \,1$, we have $N\Lb Y_{\rm min},r,b\Rb
 \ll\,1$ and the amplitude reaches the saturation region due the BFKL 
evolution (see \fig{map}-a). For such dipoles we can safely use the
 solution of \eq{SOLF} which we have discussed above.

 If the size of the dipole is large and $\tau_m \,\geq\,1$, we have
 to deal with the situation shown in \fig{map}-b, and we will discuss
 this case later.

  In the CGC/saturation approach the initial condition for the scattering
 amplitude is given 
 by the McLerran-Venugopalan (MV) formula\cite{MV}:
\beq \label{ADA1}
N_A\Lb r^2;Y; b \Rb\,\,\,=\,\,1\,\,-\,\,\,
1 - \exp\Big( - r^2 \,Q^2_s\Lb A;Y=Y_{\rm min} \Rb\Big)\,\,=\,\,1\,\,\,-\,\,e^{ - \tau_m}
\eeq
where $Q_s\Lb A;Y=Y_{\rm min} \Rb$ is the saturation momentum at the
 initial energy. For the moment we consider  the case of small
 $\tau_m$, and replace \eq{ADA1} by $N_A\Lb r^2;Y; b \Rb\,\,=\,\,\tau_m$.

For hadron-nucleus scattering
 the initial condition can be taken from the  non-perturbative 
 approach, or from the high energy phenomenology. For obvious reasons
 we have to use  phenomenology which  we have discussed 
in section II-D.  We found the value of the amplitude at $W =  
 0.576 \,GeV $   at $b=0$ is equal to 0.7 .
 To obtain the value at $Y  = Y_{\rm min}$, 
 we need to know the
 energy dependence of the amplitude,
  which  can be  obtained from  high energy
 phenomenology.  In most attempts to build such a phenomenology,
 the behaviour of the cross section with energy are assumed
 to be Reggeon-like,    $A\,\propto \,s^\Delta$ with
 $\Delta \geq 0.14$. In our own approach\cite{GLMMOD}
 the value of $\Delta \approx 0.25$, but even a value of $\Delta = 0.14$ 
 leads to the amplitude at $Y = Y_{\rm min}$ equal to  0.12 
(0.3 for our model). Therefore, it appears reasonable to assume
 that we have a situation which is shown in \fig{map}-a.  However,
 we will also consider the alternative situation which is related
 to \fig{map}-b.
  
  \begin{boldmath}
  \subsection{Processes with different multiplicities of produced 
gluons for $\tau_m \ll 1$}
  \end{boldmath}
  Rewriting \eq{SOLF} in the form
  
  \beq \label{SOLF1}
\frac{1}{\sqrt{2}}\,\int^\phi_{\phi_0}\frac{d \phi'}{\sqrt{ \phi' \,-\,1+ \,e^{-\phi'}}}\,\,\,\,+\,\,\ln \phi_0\,\,\,=\, \,\ln \phi_0 \,+\,\h\zeta\,\,=\,\,\ln\Lb \phi_0 e^{\h \zeta}\Rb
\eeq
  we can find the solution to the equation as function of $\phi_0\,
 e^{\h \zeta}$ :  $\phi\Lb \phi_0\, e^{\h \zeta}\Rb$. Practically, the 
left hand
 side of \eq{SOLF1}  does not depend on the value of $\phi_0$.
  
\begin{figure}[ht]
   \centering
  \leavevmode     
      \includegraphics[width=10cm]{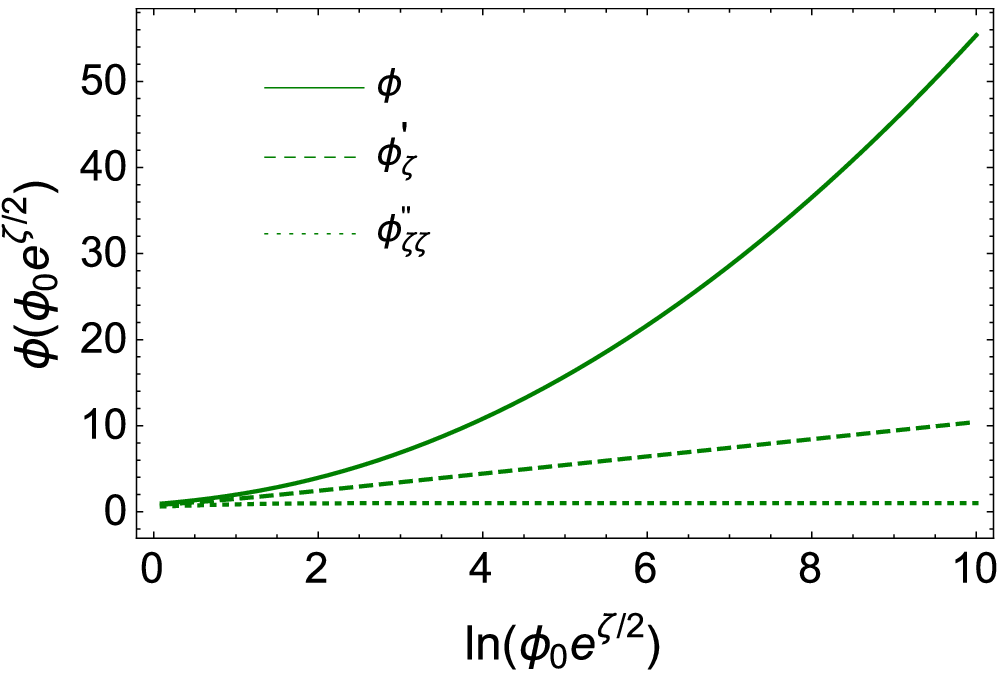}  
  \caption{The solution to \eq{SOLF1} for functions $\phi\Lb \zeta\Rb$,
 $d \phi\Lb \zeta\Rb/d \zeta$ and $d^2 \phi\Lb \zeta\Rb/d\zeta^2$.}
\label{phi}
 \end{figure}

  This function is shown in \fig{phi}.  
  
  We suggest  using the following expression for the amplitude $N\Lb 
\gamma,
 \bar{\gamma},\gamma_{\rm in}; \zeta\Rb$
  \beq \label{SOLN2}
  N\Lb \gamma, \bar{\gamma},\gamma_{\rm in}; \zeta\Rb\,\,=\,\,1 \,\,- \,\,e^{-\,\phi\Lb \gamma e^{\h \zeta}\Rb}\,\,-\,\,e^{-\,\phi\Lb\bar{ \gamma }e^{\h \zeta}\Rb}\,\,
  +\,\,e^{-\,\phi\Lb \Lb \gamma  +  \bar{\gamma}\,-\,\gamma_{\rm in}\Rb e^{\h \zeta}\Rb} 
  \eeq   

\begin{figure}[ht]
   \centering
  \leavevmode    
  \begin{tabular}{c c c} 
      \includegraphics[width=6.5cm,height=4cm]{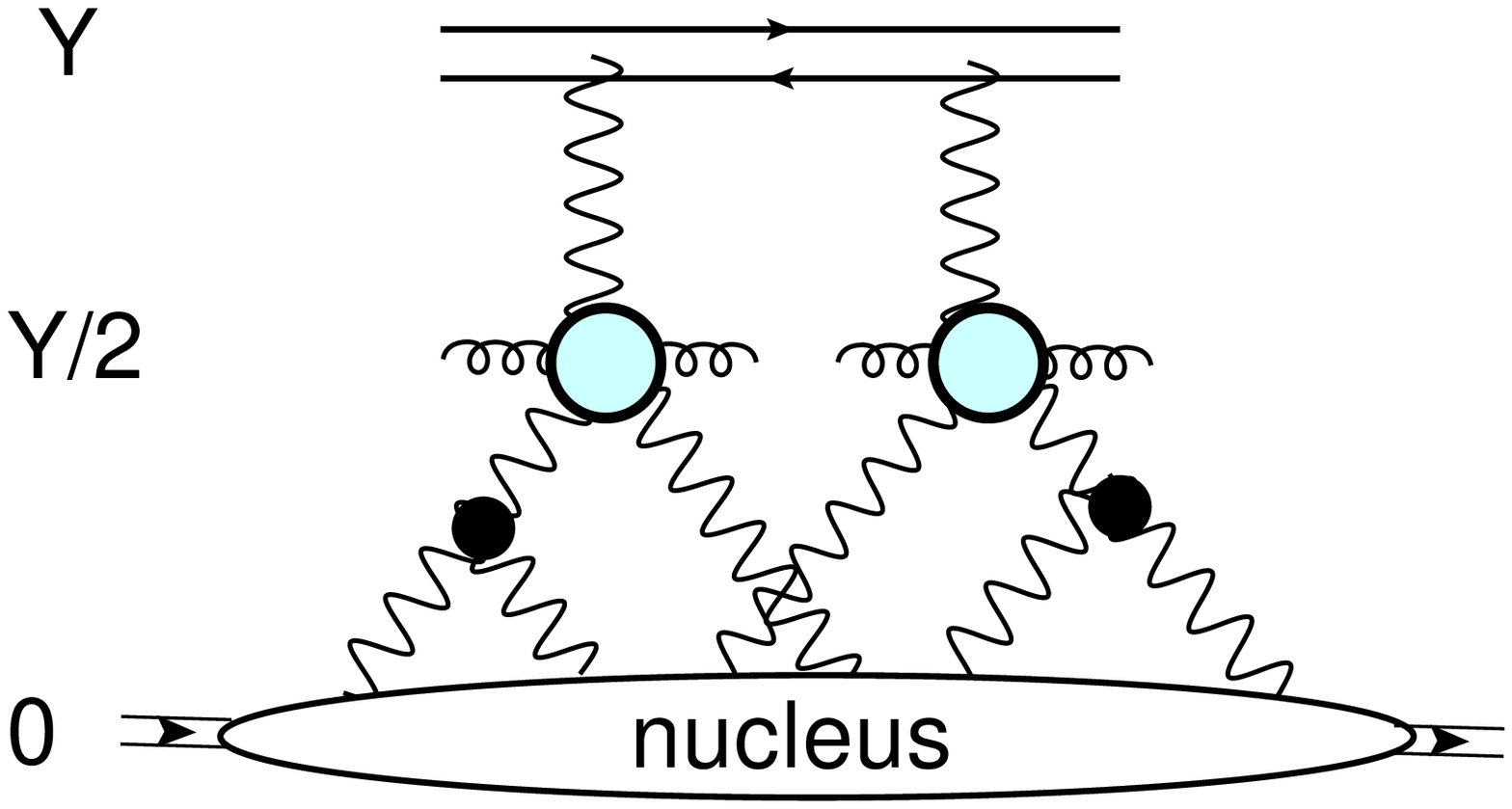}  &~~~~~~~~~~~&      \includegraphics[width=6cm]{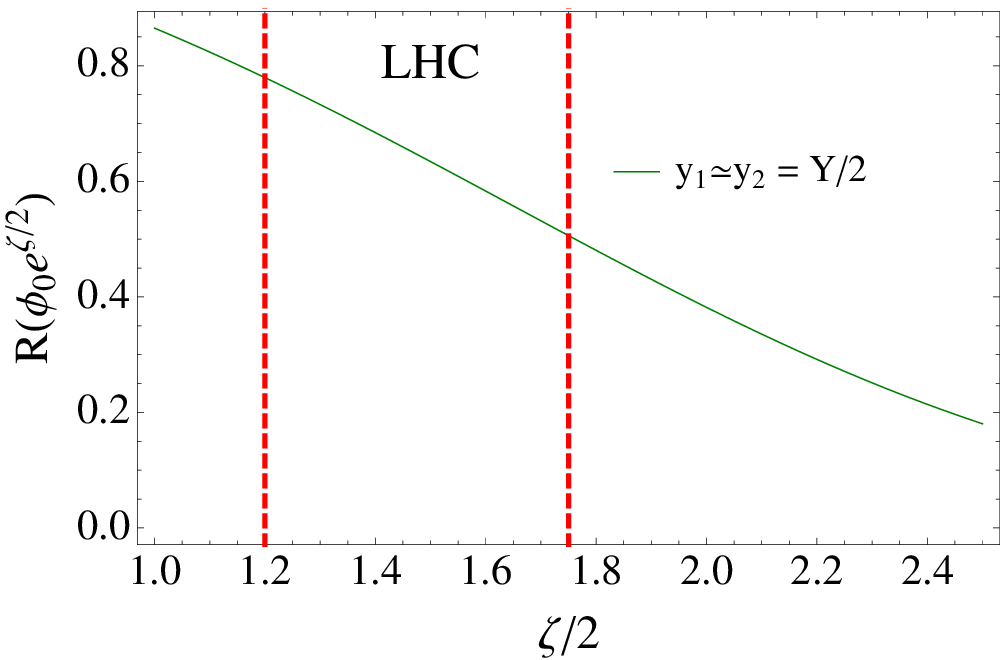} \\
      \fig{bec}-a &  & \fig{bec}-b\\ 
      \end{tabular}
       \caption{ \fig{bec}-a:  the Mueller diagram for the Bose-Einstein
 correlation function with $y_1=y_2= \h Y$.  \fig{bec}-b: the suppression
 factor $R$ of \protect\eq{RR}
       as function of $ \zeta$. The vertical dotted lines show the LHC
 kinematic region.}
\label{bec}
 \end{figure}

From  \eq{SOLN2} for the amplitude we can estimate the central diffraction
 production with multiplicity $n < 2 \bar{n}$ using \eq{GFCD}. We can use
  simplifications, as  it turns out that 
$\phi'_\zeta\,\gg\,\phi''_{\zeta
 \zeta}$ (see \fig{phi}). The cross section for central diffraction with
 multiplicity $n \,<\,2 \bar{n}$ is equal to $\sigma^{\rm CD}_0 + \sigma^{\rm CD}_1$ and can be estimated as follows:
\beq \label{SOLN3}
\sigma^{\rm CD}_0 + \sigma^{\rm CD}_1\,\,=\,\,\frac{1}{4}\Gamma^2\Lb 2 \pom
 \to  2 G\Rb\Bigg\{  \phi'^{\,2}_{\zeta}\,+\,\frac{\gamma_{\rm in}}{2
 \gamma}\,\phi'^{\,3}_{\zeta}\Bigg\} \,e^{ - \,\phi\Lb 2\,\gamma e^{\h
 \zeta}\Rb}\,\,=\,\,\,\frac{1}{4}\Gamma^2\Lb 2 \pom \to  2 G\Rb\Bigg\{
  \phi'^{\,2}_{\zeta}\,+\,\,\phi'^{ \,3}_{\zeta}\Bigg\} \,e^{ - 
\,\phi\Lb 2\,\gamma e^{\h \zeta}\Rb} 
\eeq
We assume that  $\gamma_{\rm in }  = 2 \gamma$, since  the amplitude
 in the perturbative QCD region is due to the exchange of the BFKL
 Pomeron, which has this property. The Bose-Einstein correlation can
 be evaluated using the diagram of \fig{bec}.

The contribution of this diagram  has the form\cite{AGK2}:
\beq \label{SOLN4}
\sigma^{\rm BE}\,\,=\,\,\frac{1}{N^2_c - 1} \Gamma^2_G\,\Lb \phi_0\,e^{\h ( \zeta - \h \zeta)}\Rb^2\,\Lb 1 \,-\,e^{-\,2\,\phi\Lb \phi_0 e^{\frac{1}{4} \zeta}\Rb}\Rb^2
\eeq
Finally the suppression factor $R$ is equal to
\beq \label{RR}
R\Lb \phi_0 e^{\h \zeta}\Rb\,\,=\,\,\frac{\sigma^{\rm CD}_0 + \sigma^{\rm CD}_1}{\sigma^{\rm BE}}\,\,=\,\,\Bigg\{  \phi'^{\,2}_{\zeta}\,+\,\,\phi'^{\,3}_{\zeta}\Bigg\} \,e^{ - \,\phi\Lb 2\,\phi_0 e^{\h \zeta}\Rb}\Bigg{/}\Lb\phi_0 e^{\h ( \zeta - \h \zeta)}\Rb^2\,\Lb 1 \,-\,e^{-\,2\,\phi\Lb \phi_0 e^{\frac{1}{4} \zeta}\Rb}\Rb^2
\eeq

In   \eq{RR} we did not fix the value of $y_1 \approx y_2$. The  curve
 in \fig{bec}-b is calculated for $y_1 \approx y_2 = \h Y$. The difference
 occurs since  for these estimates the argument $\phi_0 \exp\Lb \h 
\zeta\Rb$ 
of  $\phi$ in $\sigma^{\rm CD}_0 + \sigma^{\rm CD}_1$ should be replaced by
 $\phi\Lb\ln\Lb \phi\Lb \phi_0 \exp\Lb \frac{1}{4} \zeta\Rb\Rb\Rb + \frac{1}{4}
 \zeta\Rb$. For calculating the value of $ \phi_0 e^{\h \zeta}$ in this
 region we use
the energy dependence of the saturation scale
 $Q^2_s\Lb Y\Rb \propto \exp\Lb \lambda Y\Rb$
 from Ref.\cite{DKLN} ($\lambda = 0.204$) and
 value of $\phi_0 \approx\,0.3$. Note that  the
 ratio decreases at large values of $ \phi_0 e^{\h \zeta}$,
 but gives a sufficiently large  value $R \approx 0.5 - 0.75$
 in the LHC kinematic region. 

  \subsection{Estimates for proton-nucleus scattering}
  
  In this section we would like to make estimates of the damping
 factor $R$  for  proton-nucleus scattering using the same approach 
as in the Schwimmer model (section III-D  and \fig{r}). For these estimates
 we replace $\phi_0 \exp\Lb - \h \zeta\Rb$ by $\phi_0 \,S_A\Lb b \Rb 
 \exp\Lb \h \zeta\Rb$, and assumed that $\phi_0 S_A\Lb b = 0\Rb = 1/3$.
 For every value of $\zeta$, $b$ could be so large that the scattering 
amplitude becomes small and the exchange of two BFKL Pomerons, give the 
only contribution.  We found the solution of the equation $\phi_0 S_A\Lb
 b = 0\Rb  \,=\,\phi_0 S_A\Lb b_{\rm max}(\zeta)\Rb\,\exp\Lb \h \zeta\Rb$
   and replace the numerator  of \eq{RR} by the following integral:
  \beq \label{PA1}
\int^{b_{\rm max}(\zeta)}\!\!\!\! d^2 b  \Bigg\{  \phi'^{\,2}_{\zeta}\Lb 
 \phi_0 \,S_A\Lb b \Rb e^{\h \zeta}\Rb\,+\,\,\phi'^{\,3}_{\zeta} \Lb \phi_0 \,S_A\Lb b \Rb e^{ \h \zeta}\Rb\Bigg\} \,e^{ - \,\phi\Lb 2\,\phi_0 \,S_A\Lb b \Rb e^{ \h \zeta}\Rb}  \,\,+\,\,\int_{b_{\rm max}(\zeta)}\!\!\!\! d^2 b\, \Lb  \phi_0 \,S_A\Lb b \Rb\,e^{ \h \zeta} \Rb^2
\eeq
  The dominator has the form:
 \beq \label{PA2}
  \int^{b_{\rm max}(\zeta)}\!\!\!\! d^2 b\,\Lb\phi_0\,S_A\Lb b \Rb  e^{\h ( \zeta - \h \zeta)}\Rb^2 \Lb 1 \,-\,e^{-\,2\,\phi\Lb \phi_0 \,S_A\Lb b \Rb e^{\frac{1}{4} \zeta}\Rb}\Rb^2  \,\,+\,\,\int_{b_{\rm max}(\zeta)}\!\!\!\! d^2 b\, \Lb  \phi_0 \,S_A\Lb b \Rb\,e^{ \h \zeta} \Rb^2
  \eeq  
  
\begin{figure}[ht]
   \centering
  \leavevmode    
  \begin{tabular}{c c c} 
      \includegraphics[width=7cm]{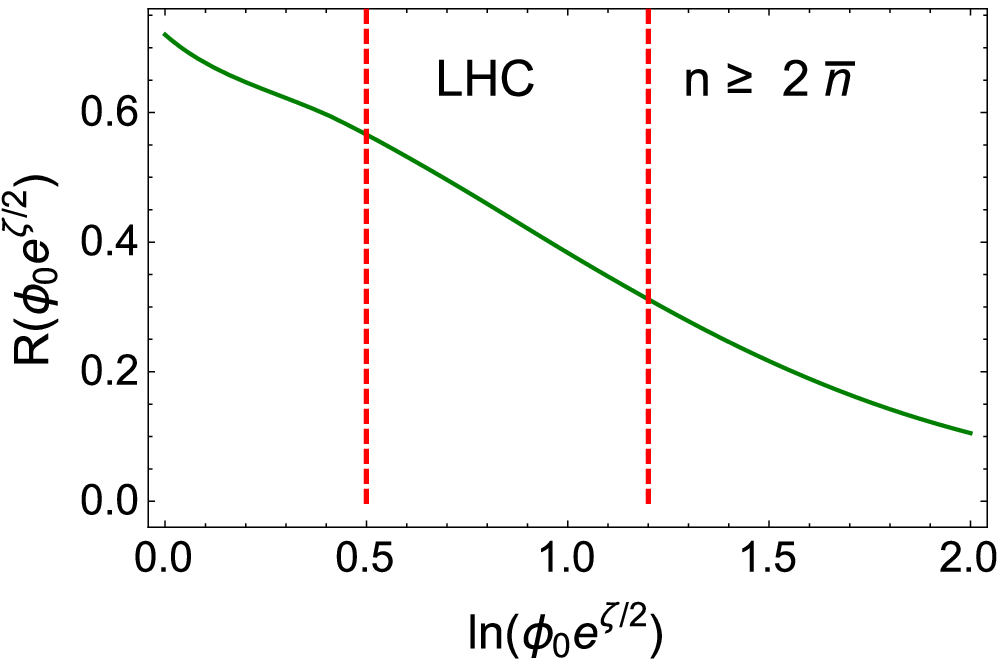}  &~~~~~~~~~~~&      \includegraphics[width=7cm]{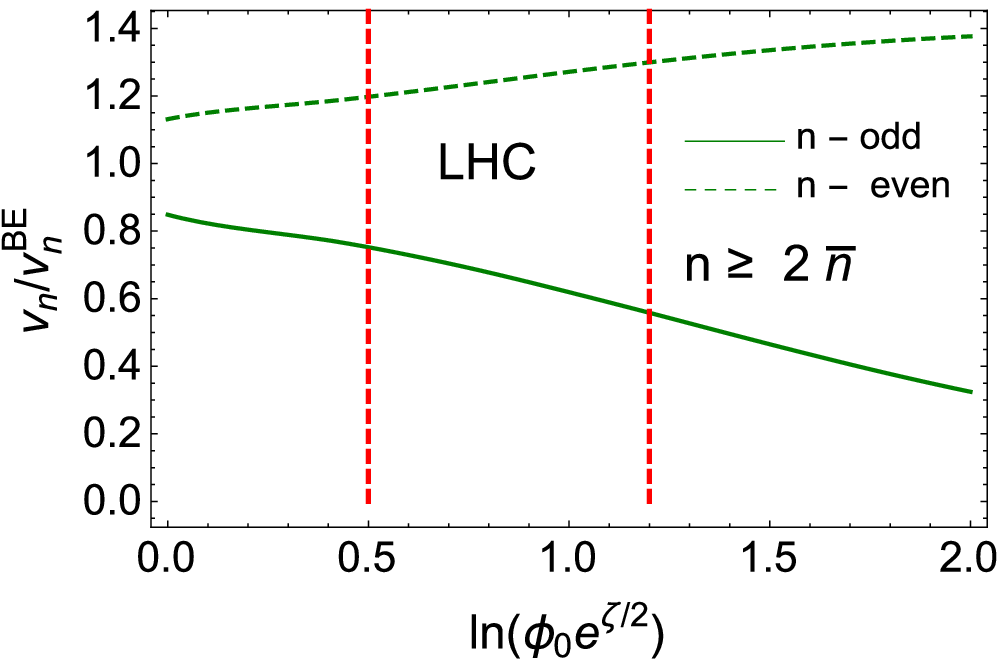} \\
      \fig{becpa}-a &  & \fig{becpa}-b\\ 
      \end{tabular}
       \caption{ For $n \geq 2 \bar{n}$ ratio $R$ (\fig{becpa}-a)  and
 the ratio $v_n/v^{\rm BE}_n$ (\fig{becpa}-b and \eq{SMV}) versus
 $\ln\Lb \phi_0 e^{\h \zeta}\Rb$. The vertical dotted lines show
 the LHC kinematic region.}
\label{becpa}
 \end{figure}

  Using \eq{PA1} and \eq{PA2} we evaluate the damping factor for
 an experiment with the multiplicity $n \geq 2 \bar{n}$, which  is
 shown in \fig{becpa}-a. In \fig{becpa}-b we plot the ratio of
 $v_n/v^{\rm BE}_n$ (see \eq{SMV}).
 This figure shows that   for energies less or about the LHC energy,
 an experiment with the selection of the multiplicities $n \geq 2
 \bar{n}$, does not gives an essential suppression for the odd harmonics.

  \begin{boldmath}
  \subsection{Processes with different multiplicities of produced gluons
 for $\tau_m \geq   1$.}
  \end{boldmath}
  
 We now  discuss  DIS with nuclei, for the case where
 $\tau_m \,\,=\,\,r^2 \,Q^2_s\Lb A;Y=Y_{\rm min} \Rb\,\,\geq\,\,1$,
 which is shown in \fig{map}-b. 

The general MV formula of 
\eq{ADA1} can be translated into the boundary conditions for $\phi$
 on the line $Y=Y_{\rm min}$ ($\xi_s=0$, see \fig{map}-b  and 
\eq{INCK1}) that has the following form:
\beq \label{BC}
\phi\Lb \xi_s=0; \xi\Rb \,\,=\,\,\phi_0 e^{\xi}
\eeq
For further discussion,  we introduce the saturation scale at 
 $Y_{\rm min}$ in a such way that $\xi \,=\,\ln\Lb r^2 \,Q^2_s\Lb A;
 Y_{\rm min}; b\Rb\Rb$ and \eq{BC} give the initial condition at $\phi_0 = 1$.

One of the general features of solution of \eq{SOLF1}, is the increase
 of $\phi$ in the saturation region (see \fig{phi}). Consequently, only
 in the vicinity of the critical line do we need to keep
 term $\exp\Lb - \phi \Rb$ in \eq{EQ}. Actually, for $\phi_0=1$, this 
term is not very large  even  at $\zeta=0$, since,
  in our estimate we are dealing with $ \gamma + \bar{\gamma}
 = 2 \phi_0  \approx 2$.

 Inside  the saturation region we can neglect this term reducing
 the equation to the simple one, namely,
\beq \label{EQC4}
\phi_{\xi_s, \xi}\,\,=\,\,\frac{1}{4};\,\,\,\,\,\mbox{or}\,\,\,\,\,\,\frac{\partial^2 \phi}{\partial t^2} \,\,-\,\,\frac{\partial^2 \phi}{\partial x^2}\,\,=\,\,\frac{1}{4}
\eeq
with the initial and boundary conditions of \eq{IC} and \eq{BC}, respectively.

It is well known that the solution of this equation is different for
 $t = \zeta\, < \,x\;( \xi <0)  $ and $t = \zeta  > x\;(\xi  > 0
 )$\cite{POL}.  For $ t = \zeta  < x \;(\xi <0)$ the solution is
 not affected by the boundary conditions, and it has the form
\beq \label{SOL11}
\phi _1\Lb z \Rb \,\,=\,\,\frac{1}{8} \zeta^2\,\,+\,\,\h\Lb e^{\phi_0}
 - 1\Rb \,\zeta\,\,+\,\,\phi_0
\eeq
Note, that for  $\phi_0$       not small, the initial condition 
of \eq{INCK1}
  reads as follows
\beq \label{INCS}
\phi\Lb \zeta = 0\Rb \,=\,\phi_0; ~~~~~~~~~~~~~~\frac{d\,\phi\Lb 
\zeta\Rb}{d\,\zeta}\,\,=\,\h\Lb e^{\phi_0}\, - \,1 \Rb;
\eeq
 The general solution to \eq{EQC4} has the form:
\beq \label{GSOL1}
\phi \Lb \xi_s, \xi \Rb \,\,=\,\frac{1}{4}\xi_s\,\xi\, +\, F_1\Lb \xi_s\Rb
 \,+\,F_2 \Lb \xi\Rb
\eeq
 Using the
 restriction from \eq{IC},
 the solution of \eq{EQC4} can be obtained from \eq{GSOL1}.
For $t  = \zeta  >  x\;( \xi > 0 )$ we need to take into account the
 boundary condition of \eq{BC}. Using the general solution in the
 form of \eq{GSOL1}, and the matching condition on the line $\xi=0$

\beq \label{MC}
\phi_1\Lb \xi = 0\Rb \,=\,\phi_2 \Lb\xi =0\Rb
\eeq
simultaneously with the boundary conditions that have the form
\beq \label{BC1}
\phi_2 \Lb\xi_s = 0 \Rb\,\,=\,\,\phi_0 e^{\xi}
\eeq
we obtain the following solution for $\xi > 0 $
\beq \label{GSOL2}
\phi_2\Lb z, \xi\Rb\,\,=\zeta^2/8\,-\,\xi^2/8 \,+\,\phi_0\,e^{\xi}\,+\,\h \Lb e^{\phi_0}\,- \,1\Rb\, \,\xi_s
\eeq
Therefore, the solution to the simplified \eq{EQC4} has
 the following form
\bea \label{SOLEQC}
\phi\Lb \zeta, \xi \Rb\,\,=\,\,\left\{\begin{array}{l}\,\,\,\phi_1\Lb \zeta \Rb\,\,\,\,\,\mbox{for}\,\,\,\xi\leq \,0\,;\\ \\
\,\,\,\phi_2\Lb \zeta, \xi\Rb
\,\,\,\,\,\mbox{for}\,\,\,\xi\,>\,0\,; \end{array}
\right.
\eea
For the solution of the general \eq{EQ} we have
\bea \label{SOLEQXIXIS}
\phi\Lb \xi_s,\, \xi;\, \Rb\,\,=\,\,\left\{\begin{array}{l}\,\,\,\phi\Lb \zeta ; \,\eq{SOLF1}  \Rb\,\,\,\,\,\mbox{for}\,\,\,\xi\leq \,0\,;\\ \\
\,\, \,\phi\Lb \xi_s + \xi; \,\eq{SOLF1}\Rb \,-\,\phi\Lb  \xi; \,\eq{SOLF1}\Rb\,+\,\phi_0\,e^{\xi}
\,\,\,\,\,\mbox{for}\,\,\,\xi\,>\,0\,; \end{array}
\right.
\eea
where $\phi\Lb \zeta ; \,\eq{SOLF1}  \Rb$ is the solution to
 \eq{SOLF1} with the initial condition of \eq{INCS}.

The solution of \eq{SOLEQXIXIS} satisfies the initial and boundary 
conditions
 of \eq{MC} and \eq{BC1}, but for $\xi > 0$ this solution leads to the
 equation
\beq \label{EQAPP}
\frac{\partial^2 \phi\Lb \xi_s, \xi\Rb}{ \partial \xi_s\,\partial \xi}\,\,=\,\,\frac{1}{4}\Big( 1\,-\,e^{ - \phi\Lb \xi_s + \xi | \eq{SOLF1}\Rb}\Big)\,\neq\,\frac{1}{4}\Big( 1\,-\,e^{ - \phi\Lb \xi_s , \xi \Rb}\Big)\eeq

\begin{figure}[ht]
   \centering
  \leavevmode    
      \includegraphics[width=8cm]{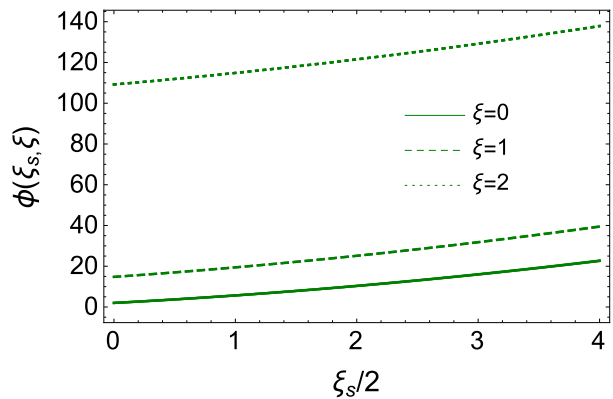} 
             \caption{ Comparison of the solutions of \eq{SOLEQXIXIS}
 and of \eq{GSOL2} with the exact numerical solutions which satisfies the
 initial and boundary conditions of \eq{MC}
            and \eq{BC1}. The solutions with the same $\xi$ are marked by
 the same type of lines. One can see that they all coincide for
 $\h \xi_s \leq 4$. The value of $\phi_0$ was taken to be  $2$ as
 follows from \eq{SOLN2N}.}
\label{comp}
 \end{figure}

 We found that this simple solution approaches
  the solution for the equation which we  found numerically
 solving \eq{EQ}, with the initial and boundary conditions of \eq{MC}
 and \eq{BC1}. In \fig{comp} we compare the solutions of
 \eq{SOLEQXIXIS}, \eq{GSOL2}  and the numerical solution
 for different values of $ \xi$.  The difference
 for $\xi \leq 8$ is not large, and \eq{SOLEQXIXIS} can be used for
 obtaining estimates. It should be noted that the simple solution of
 \eq{GSOL2} provides a good approximation of the numerical solution
 for $\gamma \sim \phi_o \approx 2$, which we need to estimate the
 damping factor.

 The  solution of \eq{SOLEQXIXIS} does not show 
 geometric scaling behavior, and the solution of \eq{EQ} depends both
 on  $\zeta  = \xi_s + \xi$ and $\xi$.

We need to generalize \eq{SOLN2} replacing $\phi$ in this formula
 by \eq{SOLEQXIXIS} which results in the following expression
\beq \label{SOLN2N}
\phi\Lb \gamma + \bar{\gamma}  - \gamma_{\rm in},\xi_s,\xi\Rb\,\,=\,\,\phi\Lb \ln\Lb \gamma + \bar{\gamma} - \gamma_{\rm in}\Rb + \h \xi_s +\h \xi ; \eq{SOLF1}\Rb \,-\,\phi\Lb \ln\Lb \gamma + \bar{\gamma} - \gamma_{\rm in}\Rb + \h \xi_s  ; \eq{SOLF1}\Rb+\,\phi_0\,e^{\xi}
\eeq

Using \eq{SOLN2N} we can calculate

\beq \label{SOLN3N}
\sigma^{\rm CD}_0 + \sigma^{\rm CD}_1\,\,=\,\,\frac{1}{4}\Gamma^2\Lb 2 \pom \to  2 G\Rb\Bigg\{ \gamma \bar{\gamma}\, \phi'^{\,2}_{\gamma}\,\Lb \gamma + \bar{\gamma} ,\xi_s,\xi\Rb\,\,+\,\,\gamma\, \bar{\gamma}\,\gamma_{\rm in}\, \phi'^{\,3}_{\gamma}\,\Lb \gamma + \bar{\gamma} ,\xi_s,\xi\Rb\Bigg\}\,e^{ - \phi\Lb \gamma + \bar{\gamma}  ,\xi_s,\xi\Rb}\eeq

\begin{figure}[ht]
   \centering
  \leavevmode   
\includegraphics[width=8cm]{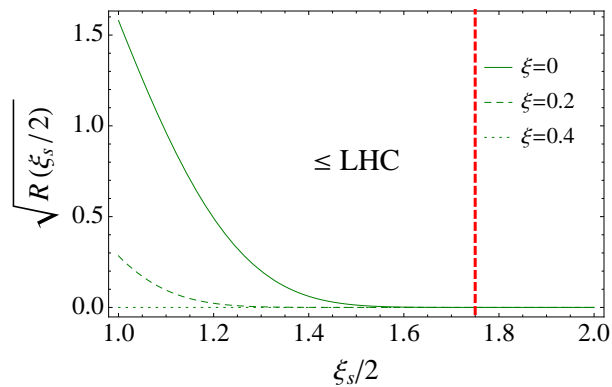}      
                 \caption{The damping factor $\sqrt{R}$ at
 different values of $\xi_s$.  For estimates for $\xi_s$ 
in the LHC kinematic region, we use $Q^2_s \propto (1/x)^\lambda $ 
 with $\lambda \approx 0.2$\cite{DKLN}.  The damping factor is plotted
 for $y_1 \approx\,y_2= \h Y$.}
\label{rxi}
 \end{figure}

  In \fig{comp} we show that the solution of \eq{SOLEQXIXIS}
depends on the value of $\xi$. Hence, we can expect that the value of
 the damping factor $R$ will depend on
$\xi$. In   \fig{rxi} we plot this dependence  for $y_1 \,\approx\,y_2$ 
 summing over all values of $y_1$.  The damping factor
 $R$ decreases, and becomes negligible at $\xi \sim 0.4$.
 Therefore,  we see the full restoration of azimuthal angular symmetry
 $\phi \to \pi - \phi$, at $\xi \geq 0.4$.

 In section III-D we used the small experimental value of the triple
 Pomeron vertex, to show that the typical size of the nucleon is so
 small, that we can safely use the geometric scaling behaviour to
 estimate the value of the damping factor. We also know that   
the triple Pomeron experimental vertex, does not show any sizable
 dependence on the momentum transferred of interacting Pomerons.
 This  can be interpreted as the small typical radius of  
the proton-nucleus
 interaction. In all attempts  to describe the interaction of protons at
 high energy, the small size of the proton components appears in different
 ways (see Ref.\cite{GLMMOD} for example where $r_{\rm proton} 
\sim 0.2 GeV^{-1}$). For such small sizes we face the situation
 shown in \fig{map}-a, for which we have a large violation of
 $\phi \to \pi - \phi$ symmetry. 

   However, for a very dense system, where the number of sources are
  large, the value of $Q^2_s\Lb A; Y_{\rm min}; b\Rb \propto
 Q^2_s\Lb Y_{\rm min}\Rb \,S_A\Lb b \Rb \propto Q^2_s\Lb Y_{\rm min}
 \Rb  \,A^{1/3}$, we always have the situation shown in \fig{map}-b
 and we deal with the violation of the geometric scaling behaviour
 for $\xi >0$ which results in the restoration of the $\phi \to \pi
 -\phi$ azimuthal angular symmetry, even for the events with 
multiplicities $n \geq 2
 \bar{n}$. For realistic heavy nuclei (gold, lead etc.
 ), $S_A\Lb b=0\Rb \sim 2$, and we could expect a  large violation of
 this symmetry, due to the selection of  events with respect
 to their multiplicities.  However, the  estimates for $y_1
 \approx\,y_2 = \h Y$, shows that such expectations are premature
 (see \fig{rxi}).
Nevertheless, it is ought to be noted that in a real experiment we
  measure the produced gluons with rather large values of
 the transverse momenta which leads to $\xi \,<\,0$, and we have
 to deal with the damping factor in \fig{map}-b, but in the region
 where we have  geometric scaling behaviour of the scattering
 amplitude.  In this region the value of the ratio
 $R$, can be obtained from \fig{rxi}-b at $\xi =0$. One can see that in this case  we
 have a rather strong violation of $\phi \,\to\,\pi - \phi$ azimuthal 
angular symmetry.


~

 \section{Conclusions}
  In this paper we demonstrated that the selection of the events
 with different multiplicities of produced particles lead to the
 violation of $\phi \to \pi - \phi$ symmetry. We found  that for 
 DIS,  if $Q^2$ is so large that $Q^2 \,>\,Q^2_s\Lb A; Y_{\rm min};
 b\Rb$ the violation of $\phi \to \pi - \phi$ symmetry turns out to
 be so large, that we can neglect  in the 
first approximation  the existence of this symmetry. For
 such $Q^2$  our estimates show that in the case, when the events
 with multiplicities  $n \geq 2 \bar{n}$ are selected,  we do not
 expect any suppression of $v_n$ for odd $n$ for the LHC energies
or lower. 
    $\bar{n}$ is the mean multiplicity at  a given energy.  
   However, for $  Q^2 \,<\,Q^2_s\Lb A; Y_{\rm min}; b\Rb $   we
 found that  for $\xi > 0.4$, we can neglect the violation of the
 symmetry and, therefore, we expect that $v_n$ with odd $n$, are small.

  Bearing this in mind, we  claim that  the character of proton-nucleus
 scattering  depends crucially    on the size of the typical 
dipole inside the
 nucleon.  There are several phenomenological observations that
 support a rather small typical radius in the nucleon, which we have
 discussed in the previous section.

  We hope that this paper will stimulate the discussion of the angular
 correlations in the events with fixed multiplicities of produced
 particles, which  crucially influence  these correlations.

  \section{Acknowledgements}
   We thank our colleagues at Tel Aviv university and UTFSM for
 encouraging discussions.   This research was supported by the BSF grant 
  2012124, by 
   Proyecto Basal FB 0821(Chile),  Fondecyt (Chile) grant 
 1180118 and by   CONICYT grant PIA ACT1406.  
 
 ~


\begin{thebibliography}{100}
\bibitem{STAR} 
  J.~Adams {\it et al.} [STAR Collaboration],
  Phys.\ Rev.\ Lett.\  {\bf 95}, 152301 (2005),
  [nucl-ex/0501016].
\bibitem{PHENIX} 
  A.~Adare {\it et al.} [PHENIX Collaboration],
  Phys.\ Rev.\ C {\bf 78}, 014901 (2008),
  [arXiv:0801.4545 [nucl-ex]].
  \bibitem{PHOBOS}
  B.~Alver {\it et al.} [PHOBOS Collaboration],
  Phys.\ Rev.\ Lett.\  {\bf 104}, 062301 (2010),
  [arXiv:0903.2811 [nucl-ex]].
  \bibitem{STAR1} 
  B.~I.~Abelev {\it et al.} [STAR Collaboration],
  Phys.\ Rev.\ C {\bf 80}, 064912 (2009),
  [arXiv:0909.0191 [nucl-ex]].
  \bibitem{CMS} 
  V.~Khachatryan {\it et al.} [CMS Collaboration],
  JHEP {\bf 1009}, 091 (2010),
  [arXiv:1009.4122 [hep-ex]].
  \bibitem{CMS1} 
  S.~Chatrchyan {\it et al.} [CMS Collaboration],
  Phys.\ Lett.\ B {\bf 718}, 795 (2013),
  [arXiv:1210.5482 [nucl-ex]].
   \bibitem{CMS2} 
  S.~Chatrchyan {\it et al.} [CMS Collaboration],
  Phys.\ Lett.\ B {\bf 724}, 213 (2013),
  [arXiv:1305.0609 [nucl-ex]].
  \bibitem{CMS3}
  S.~Chatrchyan {\it et al.}  [CMS Collaboration],
  JHEP {\bf 1402} (2014) 088,
  [arXiv:1312.1845 [nucl-ex]]. 
  \bibitem{CMS4}
    S.~Chatrchyan {\it et al.}  [CMS Collaboration],     
  Phys.\ Rev.\ C {\bf 89} (2014) no.4,  044906;\,
  [arXiv:1310.8651 [nucl-ex]].  
   \bibitem{ALICE} 
  B.~Abelev {\it et al.} [ALICE Collaboration],
  Phys.\ Lett.\ B {\bf 719}, 29 (2013),
  [arXiv:1212.2001 [nucl-ex]].
   \bibitem{ALICE1}
  B.~B.~Abelev {\it et al.} [ALICE Collaboration],
  Phys.\ Rev.\ C {\bf 90} (2014) no.5,  054901,
  [arXiv:1406.2474 [nucl-ex]]. 
    \bibitem{ALICE2}
    Y.~Zhou [ALICE Collaboration],
  J.\ Phys.\ Conf.\ Ser.\  {\bf 509} (2014) 012029.      
  \bibitem{ALICE3} 
  B.~Abelev {\it et al.} [ALICE Collaboration],
  Phys.\ Lett.\ B {\bf 719} (2013) 29,
  [arXiv:1212.2001 [nucl-ex]]. 
  
  
  \bibitem{ATLAS}
   G.~Aad {\it et al.} [ATLAS Collaboration],
  Phys.\ Rev.\ C {\bf 90} (2014) no.4,  044906;\,
  [arXiv:1409.1792 [hep-ex]].\,\,\,
  \bibitem{ATLAS1}
  B.~Wosiek [ATLAS Collaboration],
  Annals Phys.\  {\bf 352} (2015) 117.
  \bibitem{ATLAS2}
  G.~Aad {\it et al.} [ATLAS Collaboration],
  Phys.\ Lett.\ B {\bf 725} (2013) 60,
  [arXiv:1303.2084 [hep-ex]].\, 
  \bibitem{ATLAS3}
   G.~Aad {\it et al.} [ATLAS Collaboration],
  Phys.\ Rev.\ Lett.\  {\bf 116} (2016) 172301,
  [arXiv:1509.04776 [hep-ex]].
  \bibitem{GLP}
   E.~Gotsman, E.~Levin and I.~Potashnikova,
  Eur.\ Phys.\ J.\ C {\bf 77}, no. 9, 632 (2017),
  [arXiv:1706.07617 [hep-ph]].
   \bibitem{GOLE1} 
  E.~Gotsman and E.~Levin,
  Phys.\ Rev.\ D {\bf 96}, 074011 (2017),
  [arXiv:1705.07406 [hep-ph]].
  
  \bibitem{GOLE2}
  E.~Gotsman, E.~Levin and U.~Maor,
  Eur.\ Phys.\ J.\ C {\bf 76}, no. 11, 607 (2016),
  [arXiv:1607.00594 [hep-ph]].
  
  
  
    
    
  \bibitem{KOLEB}
\newblock Y.~V. Kovchegov and E.~Levin{\em { Quantum chromodynamics at high
  energy}} Vol.~33 (Cambridge University Press, 2012).

  \bibitem{KOLU1}
   A.~Kovner and M.~Lublinsky,
 Phys.\ Rev.\ D {\bf 83}, 034017 (2011),
 [arXiv:1012.3398 [hep-ph]].
  \bibitem{KOWE}
  Y.~V.~Kovchegov and D.~E.~Wertepny,
  Nucl.\ Phys.\ A {\bf 906} (2013) 50,
  [arXiv:1212.1195 [hep-ph]];\,\,\,
  
    \bibitem{KOLUCOR}
  T.~Altinoluk, N.~Armesto, G.~Beuf, A.~Kovner and M.~Lublinsky,
  Phys.\ Lett.\ B {\bf 752} (2016) 113,
  [arXiv:1509.03223 [hep-ph]];\,\,\,
  T.~Altinoluk, N.~Armesto, G.~Beuf, A.~Kovner and M.~Lublinsky,
  Phys.\ Lett.\ B {\bf 751} (2015) 448,
  [arXiv:1503.07126 [hep-ph]].
  
\bibitem{GOLE3}
E.~Gotsman and E.~Levin,
  Phys.\ Rev.\ D {\bf 95} (2017) no.1,  014034
  [arXiv:1611.01653 [hep-ph]].
    \bibitem{KOLULAST}
  A.~Kovner, M.~Lublinsky and V.~Skokov,
  {\it ``Exploring correlations in the CGC wave function: odd azimuthal anisotropy,''}
  arXiv:1612.07790 [hep-ph].
  
   \bibitem{RAJUREV}
   K.~Dusling and R.~Venugopalan,
  Phys.\ Rev.\ D {\bf 87} (2013) no.9,  094034,
  [arXiv:1302.7018 [hep-ph]] and reference therein.
  
    \bibitem{KOLUREV} 
 A.~Kovner and M.~Lublinsky,
 Int.\ J.\ Mod.\ Phys.\ E {\bf 22}, 1330001 (2013),
 [arXiv:1211.1928 [hep-ph]] and references therein.
\bibitem{KOSK} 
  Y.~V.~Kovchegov and V.~V.~Skokov,
  {\it ``When gluons go odd: how classical gluon fields generate odd azimuthal harmonics for the two-gluon correlation function in high-energy collisions,''}
  arXiv:1802.08166 [hep-ph].
 
  
\bibitem{MUDI}
A. H. Mueller,
{\it Phys. Rev.} {\bf D2} (1970) 2963.
  \bibitem{BFKL}
V.~S. Fadin, E.~A. Kuraev and L.~N. Lipatov,
\newblock Phys. Lett. {\bf B60}, 50 (1975);\,\,\,E.~A. Kuraev, L.~N. Lipatov and V.~S. Fadin,
\newblock Sov. Phys. JETP {\bf 45}, 199 (1977),
\newblock [Zh. Eksp. Teor. Fiz.72,377(1977)];\,\,\,
I.~I. Balitsky and L.~N. Lipatov,
\newblock Sov. J. Nucl. Phys. {\bf 28}, 822 (1978),
\newblock [Yad. Fiz.28,1597(1978)].



 \bibitem{BRN}
M. A. Braun,
Phys. Lett. {\bf B632} (2006) 297 [arXiv:hep-ph/0512057]; 
Eur. Phys. J. {\bf C16} (2000) 337 [arXiv:hep-ph/0001268]; 
Phys. Lett. {\bf B483} (2000) 115 [arXiv:hep-ph/0003004]; 
Eur. Phys. J. {\bf C33} (2004) 113 [arXiv:hep-ph/0309293]; 
{\bf C6}, 321 (1999) [arXiv:hep-ph/9706373]. 
M. A. Braun and G. P. Vacca,
Eur. Phys. J. {\bf C6} (1999) 147 [arXiv:hep-ph/9711486].

\bibitem{LELULOOP}
  E.~Levin and M.~Lublinsky,
  Nucl.\ Phys.\ A {\bf 763} (2005) 172,
  [hep-ph/0501173].

  \bibitem{LMP}
  E.~Levin, J.~Miller and A.~Prygarin,
  Nucl.\ Phys.\ A {\bf 806} (2008) 245,
  [arXiv:0706.2944 [hep-ph]].
\bibitem{ACKLLS}
T.~Altinoluk, C.~Contreras, A.~Kovner, E.~Levin, M.~Lublinsky and A.~Shulkin,
  JHEP {\bf 1309} (2013) 115,
  [arXiv:1306.2794 [hep-ph]].
  
  \bibitem{AKLL}
  T.~Altinoluk, A.~Kovner, E.~Levin and M.~Lublinsky,
  JHEP {\bf 1404} (2014) 075
  doi:10.1007/JHEP04(2014)075
  [arXiv:1401.7431 [hep-ph]].
  
    \bibitem{AGK} 
  V.~A.~Abramovsky, V.~N.~Gribov and O.~V.~Kancheli,
  Yad.\ Fiz.\  {\bf 18}, 595 (1973)
  [Sov.\ J.\ Nucl.\ Phys.\  {\bf 18}, 308 (1974)].

\bibitem{AGK1}
  Y.~V.~Kovchegov,
  Phys.\ Rev.\  D {\bf 64}, 114016 (2001)
  [Erratum-ibid.\  D {\bf 68}, 039901 (2003)]
  [arXiv:hep-ph/0107256].
  \bibitem{AGK2}
  Y.~V.~Kovchegov and K.~Tuchin,
  Phys.\ Rev.\  D {\bf 65}, 074026 (2002)
  [arXiv:hep-ph/0111362].
  \bibitem{AGK3}
  J.~Jalilian-Marian and Y.~V.~Kovchegov,
  Phys.\ Rev.\  D {\bf 70}, 114017 (2004)
  [Erratum-ibid.\  D {\bf 71}, 079901 (2005)]
  [arXiv:hep-ph/0405266].

\bibitem{AGK4}
  M.~A.~Braun,
  Eur.\ Phys.\ J.\  C {\bf 48}, 501 (2006)
  [arXiv:hep-ph/0603060].

\bibitem{AGK5}
  C.~Marquet,
  Nucl.\ Phys.\  B {\bf 705}, 319 (2005)
  [arXiv:hep-ph/0409023].
\bibitem{AGK6}
  A.~Kovner and M.~Lublinsky,
  JHEP {\bf 0611}, 083 (2006)
  [arXiv:hep-ph/0609227].
  \bibitem{AGK7}
   E.~Levin and A.~Prygarin,
  Phys.\ Rev.\ C {\bf 78}, 065202 (2008),
  [arXiv:0804.4747 [hep-ph]].
      \bibitem{AGK8}
  J.~Jalilian-Marian and Y.~V.~Kovchegov,
  Phys.\ Rev.\  D {\bf 70}, 114017 (2004)
  [Erratum-ibid.\  D {\bf 71}, 079901 (2005),
  [arXiv:hep-ph/0405266].
  
  
  \bibitem{BK}
I.~Balitsky,
[arXiv:hep-ph/9509348];\,\,
{\it Phys.\ Rev.} {\bf D60}, 014020 (1999)
[arXiv:hep-ph/9812311];\,\,\,\,
Y.~V.~Kovchegov,
{\it Phys.\ Rev.}  {\bf D60}, 034008  (1999),
[arXiv:hep-ph/9901281].
        \bibitem{MUDIP}
A.~H.~Mueller,
 Nucl.\ Phys.\, B \, {\bf  415} (1994) 373;
{\it ibid}  {\bf 437} (1995) 107.
\bibitem{LELU1}
E.~Levin and M.~Lublinsky,
  Nucl.\ Phys.\ A {\bf 730} (2004) 191,
  [hep-ph/0308279].
  
  \bibitem{LELU2}
  E.~Levin and M.~Lublinsky,
  Phys.\ Lett.\ B {\bf 607} (2005) 131,
  [hep-ph/0411121].
  \bibitem{KOLE}
Y.~V.~Kovchegov and E.~Levin,
  Nucl.\ Phys.\ B {\bf 577} (2000) 221
  [hep-ph/9911523].
  
    \bibitem{BGLM}
  S.~Bondarenko, E.~Gotsman, E.~Levin and U.~Maor,
  Nucl.\ Phys.\ A {\bf 683} (2001) 649,
  [hep-ph/0001260].
  \bibitem{BKKMR}
  K.~G.~Boreskov, A.~B.~Kaidalov, V.~A.~Khoze, A.~D.~Martin and M.~G.~Ryskin,
  Eur.\ Phys.\ J.\ C {\bf 44} (2005) 523,
  [hep-ph/0506211].
  
  
  \bibitem{LEPR}
  E.~Levin and A.~Prygarin,
  Eur.\ Phys.\ J.\ C {\bf 53} (2008) 385,
  [hep-ph/0701178].
  \bibitem{KTM}
  I.A. Verdiev, O.V. Kancheli, S.G. Matinyan, A.M. Popova and K.A. Ter-Martirosyan, Sov.
Phys. JETP 19, 1148 (1964) . 
  \bibitem{FK} 
  J.~Finkelstein and K.~Kajantie,
  Phys.\ Lett.\  {\bf 26B}, 305 (1968).
  \bibitem{ANSW} 
  H.~D.~I.~Abarbanel, J.~B.~Bronzan, R.~L.~Sugar and A.~R.~White,
  Phys.\ Rept.\  {\bf 21}, 119 (1975).
  \bibitem{KMRFK}
  V.~A.~Khoze, A.~D.~Martin and M.~G.~Ryskin,
  Phys.\ Lett.\ B {\bf 780} (2018) 352
  [arXiv:1801.07065 [hep-ph]].
  \bibitem{DL}
 {\it  Phys.Lett.} {\bf B437} (1998) 408 and references therein.
 
  \bibitem{SCHW}
A. Schwimmer, Nucl. Phys. B94 (1975) 445.  

\bibitem{GLM2CH}
   E.~Gotsman, E.~Levin and U.~Maor,
  Eur.\ Phys.\ J.\ C {\bf 75} (2015) no.5,  179
  doi:10.1140/epjc/s10052-015-3399-4
  [arXiv:1502.05202 [hep-ph]].
  \bibitem{KLP}
  A.~Kormilitzin, E.~Levin and A.~Prygarin,
  Nucl.\ Phys.\ A {\bf 813}, 1 (2008)
  [arXiv:0807.3413 [hep-ph]].
  
  \bibitem{LETU}
    E.~Levin and K.~Tuchin,
  Nucl.\ Phys.\ B {\bf 573}, 833 (2000),
  [hep-ph/9908317].
  \bibitem{LEPP}
   E.~Levin,
  JHEP {\bf 1311}, 039 (2013),
  [arXiv:1308.5052 [hep-ph]].
  \bibitem{KLT} 
   A.~Kormilitzin, E.~Levin and S.~Tapia,
  Nucl.\ Phys.\ A {\bf 872} (2011) 245
  [arXiv:1106.3268 [hep-ph]].
   \bibitem{DGLAP}
 V. N. Gribov and L. N. Lipatov,  Sov. J. Nucl. Phys {\bf 15} (1972)
                438;\\
 G. Altarelli and G. Parisi, {Nucl. Phys.} {\bf B 126} (1977) 298; \\
Yu. l. Dokshitser, { Sov. Phys. JETP} {\bf 46}  (1977) 641.  

\bibitem{GSV}
E.~Iancu, K.~Itakura and L.~McLerran,
  Nucl.\ Phys.\ A {\bf 708} (2002) 327
  [hep-ph/0203137];\,\,\,A.~H.~Mueller and D.~N.~Triantafyllopoulos,
  Nucl.\ Phys.\ B {\bf 640} (2002) 331
  [hep-ph/0205167];\,\,D.~N.~Triantafyllopoulos,
  Nucl.\ Phys.\ B {\bf 648} (2003) 293
  [hep-ph/0209121].
  \bibitem{POL}
Andrei D. Polyanin and Valentin F. Zaitsev, Handbook of nonlinear Partial Differential Equations,  Chapman \& Hall/CRC, 2004.  
\bibitem{GLMMOD}
 E.~Gotsman, E.~Levin and I.~Potashnikova,
 Phys. Letters B {\bf 781}, (2018) 155 [arXiv:1712.06992 [hep-ph]];\,\,\,
  Eur.\ Phys.\ J.\ C {\bf 77} (2017) no.9,  632
  [arXiv:1706.07617 [hep-ph]];\,\,\, E.~Gotsman, E.~Levin and U.~Maor,
  Eur.\ Phys.\ J.\ C {\bf 75} (2015) no.5,  179,
  [arXiv:1502.05202 [hep-ph]].
 \bibitem{DKLN}
  A.~Dumitru, D.~E.~Kharzeev, E.~M.~Levin and Y.~Nara,
  Phys.\ Rev.\ C {\bf 85} (2012) 044920,
  [arXiv:1111.3031 [hep-ph]].
  
    \bibitem{MV}
L. McLerran and R. Venugopalan, 
Phys. Rev. {\bf D49} (1994) 2233, 3352; {\bf D50} (1994) 2225; 
{\bf D53} (1996) 458;\\ {\bf D59} (1999) 09400. 
  
  
   \end{thebibliography}
\end{document}